\documentclass[twocolumn]{aastex701}

\usepackage{comment}
\usepackage{hyperref}
\usepackage{enumitem}
\usepackage{amsmath}
\usepackage{graphicx}
\usepackage{natbib}

\begin{document}

\title{Improvements to Asteroseismic Fitting of White Dwarfs in the \textit{Gaia} Era}

\author[0009-0008-2682-7088]{Andrew H.\ Dublin}
\affiliation{Department of Physics, City University of New York Graduate Center, 365 5th Ave, New York, NY 10016, USA} 
\affiliation{Physics Department of Queens College, Queens College Science Building, 6530 Kissena Blvd B334, Queens, NY 11367, USA}
\email[show]{Andrew.Dublin85@gc.cuny.edu}  

\author[0000-0002-0656-032X]{Keaton J.\ Bell}
\affiliation{Department of Physics, City University of New York Graduate Center, 365 5th Ave, New York, NY 10016, USA} 
\affiliation{Physics Department of Queens College, Queens College Science Building, 6530 Kissena Blvd B334, Queens, NY 11367, USA}
\affiliation{Department of Astrophysics, American Museum of Natural History, Central Park West at 79th Street, New York, NY 10024, USA}
\email[]{keaton.bell@qc.cuny.edu} 

\author[0000-0002-7487-9340]{Agn{\`e}s Bischoff-Kim}
\affiliation{Penn State Wilkes-Barre, 44 University Drive, Dallas, PA 18612, USA}
\email[]{axk55@psu.edu} 

\begin{abstract}

White dwarf asteroseismology aims to constrain the interior structures of pulsating white dwarf stars by fitting observed pulsation periods to those computed for stellar models. However, there remain several obstacles for achieving reliable results with the model grid-fitting approach. We simulate data for a parameter recovery experiment in one and two dimensions (mass and effective temperature) to demonstrate improved methodologies that address challenges for characterizing degenerate asteroseismic solutions. We show that interpolating model periods onto a finer grid can adequately resolve asteroseismic solutions that would be missed with sparse model grids. Incorporating absolute magnitude from \textit{Gaia} astrometry into the statistical fitting is shown to reduce solution degeneracy. We define a seismic solution as a solution to the mode identification problem, and we show that fitting to consistent model pulsation modes can isolate each candidate solution in degenerate solution space. A criterion based on $\chi^2$ for what should be considered a reasonable fit is adopted, and we identify all combinations of model periods that meet this criterion, not only those nearest to the measured periods. Finally, we demonstrate how fitting Gaussians to probability distributions allows for the robust characterization of each candidate solution, including uncertainties. Our accurate characterization of the degenerate solution landscape is supported by the comparison of our solutions to a direct marginalization of the likelihood function. These fitting approaches can be generalized to higher dimensions, where there are more than two free parameters.

\end{abstract}

\keywords{\uat{Asteroseismology}{73} --- \uat{Astrometry}{80} --- \uat{Astronomy data analysis}{1858} --- \uat{Astrostatistics}{1882} --- \uat{White dwarf stars}{1799}
}

\section{Introduction} \label{sec:intro}
	
	White dwarf stars are the evolutionary remnants of low- to intermediate-mass ($M_{\star}\lesssim10\,M_{\odot}$) main sequence stars \citep{Ibeling2013ApJ}. They are essential to our understanding of stellar astrophysics, providing a fossil record of stellar evolution, star formation, and mass loss since the birth of our galaxy \citep{WingetKepler2008, Althaus2010Review, Corsico2019Review}. White dwarfs are high-energy astronomical laboratories in their own right, whose interior structures allow us to probe physics at extreme densities, temperatures, and pressures \citep{Althaus2010Review, Saumon2022}. A thorough understanding of white dwarfs has benefited the physics community at large, though our direct constraints on white dwarf internal structure and composition remain limited. 
	
Asteroseismology provides an opportunity to characterize the interior structures of pulsating white dwarfs by interpreting the resonant frequencies of their standing wave oscillations. White dwarf stars evolve by cooling, and they pulsate during the cooling process, exhibiting measurable, periodic brightness variations at their photospheres upon formation of a partial ionization zone \citep{WingetKepler2008}. Of the white dwarfs that pulsate, the majority are the hydrogen-atmosphere white dwarfs (DAVs), which pulsate in an ``instability strip" of effective temperatures between $10{,}600\,{\rm K} \leq T_{\text{eff}} \leq 12{,}500\,{\rm K}$ for typical $\approx0.6\,M_\odot$ white dwarfs \citep{Tremblay2015}. The helium-atmosphere white dwarfs (DBVs) pulsate between $22{,}000 \,{\rm K} \leq T_{\text{eff}} \leq 30{,}000 \,{\rm K}$ \citep{Vanderbosch2022}. 
In addition, a third class of pulsating white dwarf, remnant cores of planetary nebulae, pulsate in the ``DOV'' instability strip of around $80{,}000 \,{\rm K} \leq T_{\text{eff}} \leq 180{,}000 \,{\rm K}$ \citep{Sowicka2023}. These gravity-mode pulsations cause brightness variations at resonant frequencies that are highly sensitive to the interior structure and chemical composition of the star. These nonradial pulsation modes are interpreted to have the patterns of spherical harmonics on the surface with integer spherical degrees $\ell$, azimuthal orders $m$, and radial orders $k$ \citep[for more details, see][]{Bell2025Encyclopedia}. Due to geometric cancellation in disc-integrated light \citep{Dziembowski1977}, we only expect to detect variability from $\ell =1$ and $\ell=2$ modes. Pulsation periods for typical $\sim0.6 M_{\odot}$ white dwarfs are usually between two to twenty minutes \citep{Mukadam2006}. 

A common approach to white dwarf asteroseismology aims to find matches (``solutions") between pulsation periods calculated for stellar models and those measured from photometric light curves, a procedure known as period-by-period fitting. The models are either evolutionary, where interior structures are determined by modeling of physical processes through all phases of stellar evolution preceding the white dwarf phase \citep[e.g.,][]{Althaus2010, Romero2012, Romero2022, Althaus2022,  Romero2025}, or more flexible parameterized structure models, where chemical profiles are prescribed and can represent structures that depart from expectations from evolutionary calculations \citep{BischoffKimMontgomery2018, Giammichele2022}. 
Since computing stellar models can be computationally expensive, fitting pulsation period measurements to models is typically done using a grid of precomputed models that span the pulsational instability strip \citep{BischoffKim2014, Castanheira2018, Corsico2019, Althaus2022,  BischoffKim2023}, though some works have used other search algorithms to seek optimal models that converge to the data \citep{Metcalfe2000,Charpinet2015,Bell2019}.

The residuals obtained between the best fit models and the observed data using the period-by-period fitting approach are typically orders of magnitude larger \citep{Bell2019, Romero2022, Hall2023} than the measurement uncertainties, with residuals on the order of seconds and uncertainties on the order of tenths of seconds. Period-by-period fitting is also plagued by the issue of solution degeneracy \citep{Hall2023}, whereby multiple combinations of parameters such as mass and effective temperature yield similar numerical solutions. This is a consequence of the fact that, in general, we do not know which pulsation modes we are observing. Often times, this stymies the determination of convincing global minima for goodness-of-fit values. Past studies have utilized photometric, astrometric, and/or spectroscopic comparisons to further narrow down the possible solution space or as a consistency check \citep{Castanheira2008,Romero2012,Giammichele2022,Hall2023,BischoffKim2023,Calcaferro2024}. Often, period-by-period fitting only yields suggestive best-fit models that are insufficiently resolved in parameter space, with no indication of reliable uncertainties. The notable exception is the work of \citet{Giammichele2018, Giammichele2022}, who achieved the first seismic solutions to within the uncertainties of the measured pulsation periods. In order to obtain such a precise result, the model proposed by \citet{Giammichele2018} for the DBV pulsator KIC 08626021 required a $40\%$ more massive core and a $15\%$ more oxygen-dominant core than expected from stellar evolution models \citep{DeGeronomo2019}, challenging our current understanding of late-stage stellar evolution \citep{Timmes2018, CorsicoAlthaus2024}. 
	
	 White dwarf science has recently burgeoned since the advent of the European Space Agency's \textit{Gaia} spacecraft mission. As a global astrometric mission, \textit{Gaia} has, among its many goals, monitored the positions and brightnesses of myriad celestial objects with unprecedented accuracy \citep{GaiaDR32023}. Bright white dwarfs, due to their intrinsically low luminosities, must necessarily be close-by, which means they exhibit large parallaxes. By converting constraints on parallax distance to constraints on radius as a function of effective temperature \citep{GentileFusilloDR2}, an absolute magnitude can be determined.  \textit{Gaia} Early Data Release 3 (EDR3) parallax measurements have been used to catalog 359{,}000 high-confidence white dwarf candidates \citep{GentileFusillo2021}.  
     
     For a seismic model of interest, an absolute magnitude can be estimated; the distance required to make that model appear as faint as the observed apparent magnitude can be calculated. This ``seismic distance'' \citep{Kawaler1994, Bradley1998, Bell2019} can then be compared for accuracy to the reported \textit{Gaia} distance constrained by parallax, as an afterthought to statistical fitting. \textit{Gaia} astrometric data, in conjunction with time series photometric \textit{TESS} data, has been used to support the accurate asteroseismic fitting of DAVs \citep{Romero2019, Romero2022} and DBVs \citep{Bell2019}. \textit{Gaia} astrometric data, in tandem with spectroscopically derived measurements of surface gravity and effective temperature, has been used to select preferred models amongst degenerate asteroseismic results \citep{Uzundag2023}. This approach has provided asteroseismic constraints on mass and effective temperature of DAVs \citep{BischoffKim2023,BischoffKimBell2024} and also elucidated structural parameters and chemical profiles \citep{BischoffKimBell2024}. In other cases, however, this approach has exposed the shortcomings of certain white dwarf asteroseismic modeling efforts \citep{Lopez2021,Bognar2026} and highlighted the inaccuracies between seismic fits and spectroscopic results \citep{Romero2019}. Moreover, significant discrepancies have been shown to exist between astrometric masses and mass estimates derived from spectroscopy, photometry, and seismology for certain DAVs and myriad DBVs \citep{Calcaferro2024}. Despite the success of \citet{Giammichele2018} in achieving a precise asteroseismic fit, the \textit{Gaia} radius constraint for KIC 08626021 is discrepant with their seismic solution by more than $6\sigma$ \citep{Bell2022}. This suggests that the solution identified by \citeauthor{Giammichele2018}\ is inaccurate, potentially because of a misinterpretation of which modes were observed. These results demonstrate current limitations of asteroseismic techniques that could benefit from improved methodology, as well as the value of astrometry to support the determination of accurate seismic results.
	
	In this work, we demonstrate a statistical framework for resolving and characterizing asteroseismic solutions in a degenerate parameter space. Using a parameter recovery test with simulated data, we demonstrate how (i) interpolation of model periods over a coarse grid can help identify and resolve asteroseismic solutions. We show that (ii) incorporating \textit{Gaia} absolute magnitude directly into the period-by-period fitting helps to reduce solution degeneracy. We propose (iii) a definition for a seismic solution as a solution to the mode identification problem, and we demonstrate a revision to the quality function that allows for individual solutions to be isolated and characterized. We (iv) apply this quality function to all viable combinations of model modes rather than only considering those nearest to the measured periods, which could cause valid solutions to be missed. This enables the complete characterization of degenerate solution space with (v) robust determinations of parameter uncertainties from fitting multivariate Gaussian distributions to each individual solution. 
		
\section{Simulating Data} \label{sec:Section 2}

This work presents an asteroseismic parameter recovery test for simulated white dwarf data, with the goal of demonstrating a robust methodology for characterizing asteroseismic results. For our one- and two-dimensional parameter recovery experiments, we use the White Dwarf Evolution Code \citep[WDEC; ][]{BischoffKimMontgomery2018} for generating parameterized structure models. WDEC evolves hot polytropic models ($\sim\!100{,}000\,{\rm K}$) down to a specified effective temperature, with the models relaxed along the way in order to satisfy the equations of stellar structure. WDEC parameterizes the core oxygen profile as well as the envelope structure, and it computes the pulsation periods of dipole and quadrupole modes. 

\begin{figure}
    \centering
    \par\medskip 
    \includegraphics[width=0.45\textwidth]
    {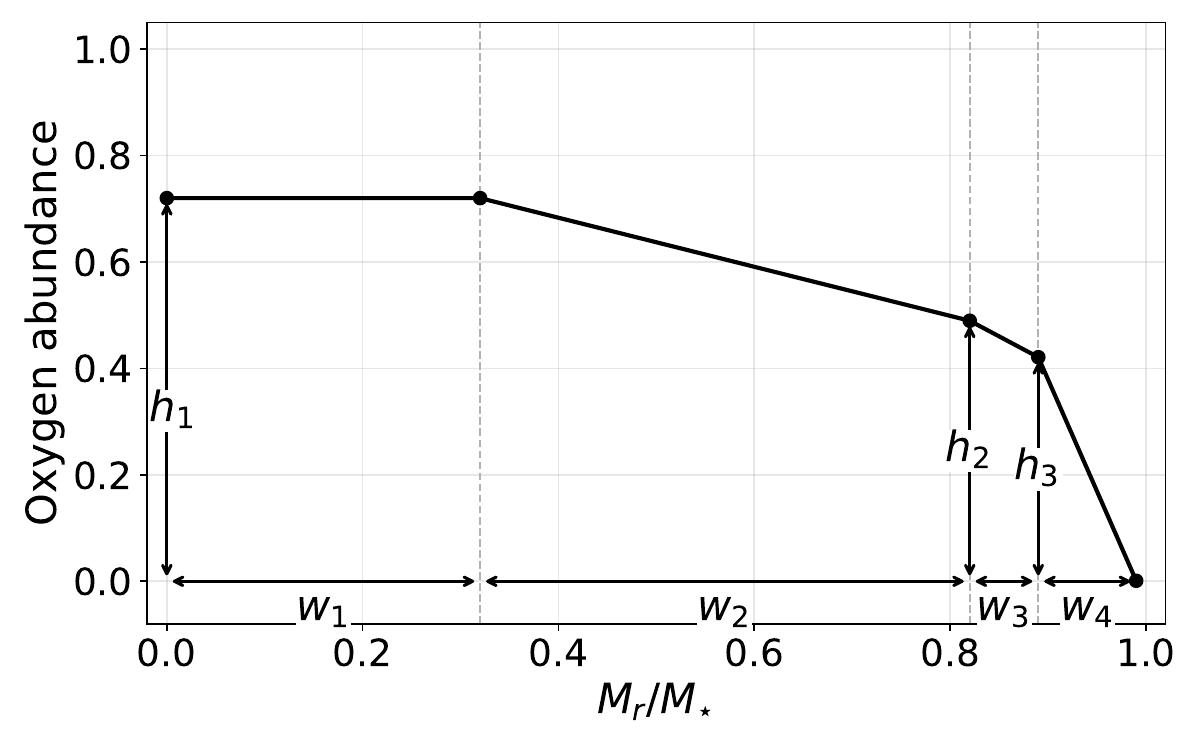}
    \caption{Input WDEC oxygen abundance profile of the core for the fiducial model described by the parameters in Table~\ref{tab:wdecparams}. The input profile gets smoothed before use in the model to produce the abundance curve shown in Figure~\ref{fig:chem_abundance}.}
    \label{fig:oxygen_abundance}
\end{figure}

\begin{figure}
    \centering
    \par\medskip
    \includegraphics[width=0.45\textwidth]
    {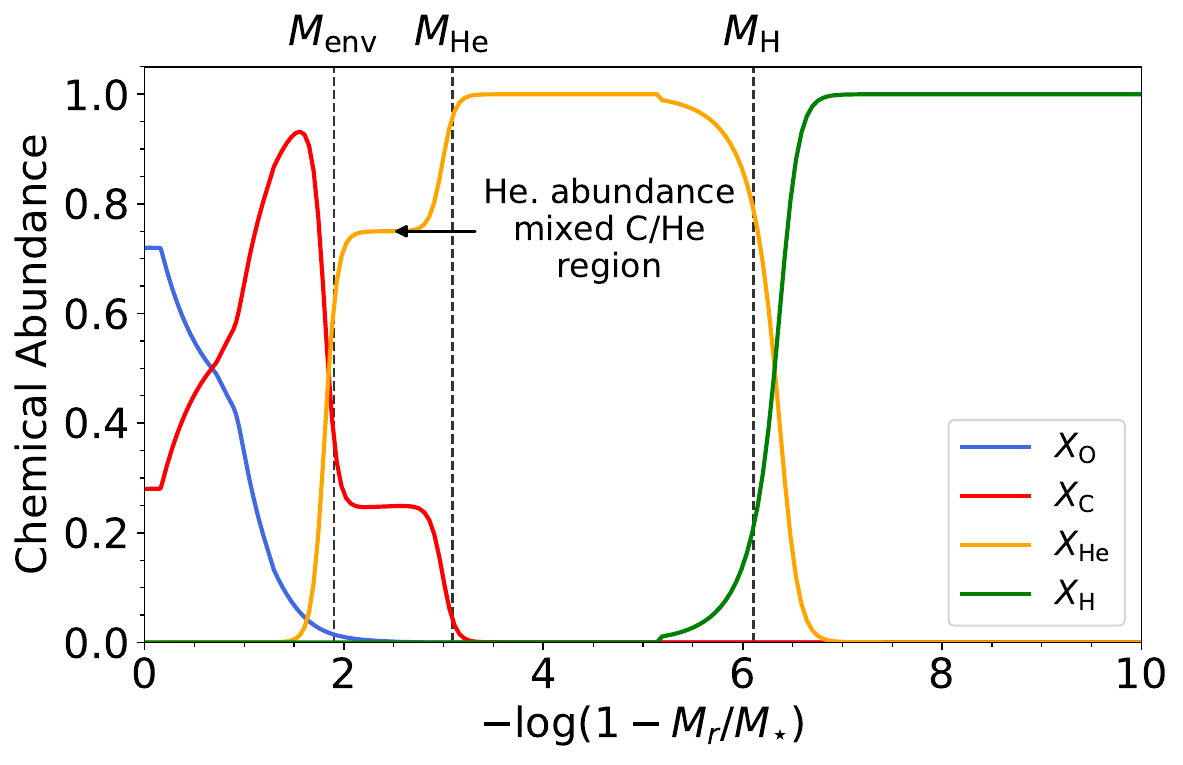}
    \caption{Final chemical abundance profile for the fiducial model described by the parameters in Table~\ref{tab:wdecparams}.}
    \label{fig:chem_abundance}
\end{figure}

The work of \cite{Yao2025} examines an application of WDEC in the asteroseismological analysis of a DAV pulsator, TIC 231277791, with ten reported independent pulsation modes. The authors use WDEC to identify two optimal models that they consider good fits to the data. To test methods for white dwarf asteroseismology, we compute a fiducial DAV model with WDEC with parameters matching their ``optimal model 2'', with effective temperature $T_{\text{eff}} = 11{,}910\,{\rm K}$ and total mass $M_\star = 0.720 $\,M$_{\odot}$. The parameters that define the interior chemical composition profile \citep[see][]{BischoffKimMontgomery2018} are given in Table~\ref{tab:wdecparams}.In Figure~\ref{fig:oxygen_abundance} we display the initial oxygen abundance profile of the core defined by parameters $h_i$ and $w_i$. Figure~\ref{fig:chem_abundance} shows the full chemical profile of the fiducial model after the core oxygen profile has been smoothed. We will aim to recover the global parameters of $T_{\text{eff}}$ and $M_\star$ in one- and two-dimensional grid-fitting experiments using models with interior structure fixed to the values in Table~\ref{tab:wdecparams}. This reference model was chosen arbitrarily from the recent literature and the specific choice should not significantly impact our proceeding demonstration of improved methodologies for asteroseismic fitting. 

\begin{table}
\centering
\begin{tabular}{||c c||}
\hline
Parameter & Value \\ [0.5ex]
\hline\hline

$T_\text{eff}$ & 11,910 ${\rm K}$ \\
Mass & 0.720 $M_{\odot}$ \\
$-\log(M_\text{env}/M_{\star})$ & 1.90 \\
$-\log(M_\text{He}/M_{\star})$ & 3.09 \\
$-\log(M_{\text{H}}/M_{\star})$ & 6.11 \\
He. abund. mixed C/He/H region & 0.75 \\
diff. coeff. He. base env. & 16 \\
diff. coeff. He. base pure He. & 16 \\
\text{MLT}/$\alpha$ & 0.65 \\
$h_{1}\times 100$ & 72.0 \\
$h_{2}\times 100$ & 48.96 \\
$h_{3}\times 100$ & 42.11\\
$w_{1}\times 100$ & 32.0 \\
$w_{2}\times 100$ & 50.0 \\
$w_{3}\times 100$ & 7.0 \\
$w_{4}\times 100$ & 10.0 \\

\hline
\end{tabular}
 \caption{Parameters of the WDEC model used to generate the simulated data analyzed in our parameter recovery test. Most parameters were chosen to match  ``optimal model 2'' described by \cite{Yao2025}. The values for the two diffusion coefficients, as well as the mixing length $\alpha$ coefficient, were selected from the fiducial values in Table 2 from \cite{BischoffKim2023August}. The value of $w_{4}$ was chosen to be the minimum of the range of values listed for $w_{3}$. The values $h_{2}$ and $h_{3}$ satisfy $h_{2} \times 100 = 0.68 \cdot h_{1} \times 100$, $h_{3} \times 100 = 0.86 \cdot h_{2} \times 100$, respectively. See \cite{Bischoff-Kim2018c} for a description of the envelope and core parameters.}\label{tab:wdecparams}
\end{table}

We select three axisymmetric ($m=0$) modes randomly from the periods computed for the true model, and we add 1.0-second Gaussian noise to each. The work of \cite{Giammichele2022} considers three potential global sources of model uncertainties for structural models and find that the presence of Neon-22 introduces the largest uncertainty (0.43 seconds) in their static models. Our noise level of 1.0 second is chosen carefully to exceed this value, since additional sources of systematic error may still be present. These three simulated periods were then treated as the `observed' periods used in the asteroseismic fitting, with $P_{1} = 345.54$\,s $(\ell=1, k=4)$, $P_{2} = 615.09$\,s $(\ell=2, k=22)$, and $P_{3} = 1092.09$\,s $(\ell=2, k=41)$, each with the associated 1.0-s statistical error. We choose to ``observe'' three periods specifically to slightly exceed the number of free parameters in this two-dimensional parameter variation experiment. In a full asteroseismic analysis that varies a greater number of parameters associated with the stellar interior structure, typically more observed periods are needed to obtain useful constraints. Even with more observed periods than parameters, solution degeneracy arises because each observed period can be interpreted as corresponding to a variety of physical pulsation modes in the models.

We use synthetic photometry \citep{ Bergeron1995, Holberg2006} available\footnote{\url{http://www.astro.umontreal.ca/~bergeron/CoolingModels}} for DA white dwarf atmosphere models \citep{Tremblay2011} to determine the absolute magnitude of our WDEC model in the \textit{Gaia} DR3 $G$ band \citep{Riello2021}. Bilinear interpolation of their model grid yields a magnitude and radius corresponding to the effective temperature and mass of our WDEC model. The interpolated $G$-band magnitude represents the emergent flux spectrum computed for this stellar atmosphere, but for white dwarf models with different interior structures \citep{Bedard2020}---and therefore radii---than our WDEC model. We rescale the model magnitude from the interpolated value ($M_{G\text{,interpolated}}$) to account for the difference in radius between our WDEC model ($R_{\text{WDEC}}$) and the radius interpolated for the atmosphere model ($R_{\text{interpolated}}$) as 
	\begin{equation}
	M_{G,{\rm model}} = M_{G,\text{interpolated}} -2.5 \log \left( \frac{R_{\text{WDEC}}}{R_\text{interpolated}}\right)^{2}.
    \label{eq:interpolation}
	\end{equation}
For the WDEC model with parameters given in Table~\ref{tab:wdecparams}, we compute an absolute $G$-band magnitude of $M_G = 12.166$\,mag. We add random Gaussian noise with $\sigma_{\rm mag} = 0.05$ mag representative of typical uncertainties from the \textit{Gaia} EDR3 catalogue of white dwarf stars \citep{GentileFusillo2} with colors consistent with DAVs and whose apparent magnitudes ($G\sim17$) are typical of DAVs observed by TESS \citep{Romero2019}. This yields a ``measured'' absolute magnitude for our WDEC model of $M_G=12.202\pm0.05$\, mag.

\section{Asteroseismic Characterization} \label{sec: Section 3} 

Here we present and test our improvements to the fitting methodology that enables us to identify and characterize seismic solutions. For our simulated data, we aim to recover the physical parameters of the model in Table~\ref{tab:wdecparams}, along with any other degenerate solutions that can explain the data. 

As a baseline for fitting asteroseismic models to measured periods, we adopt $\chi^{2}$ as our metric for goodness of fit, which is often used as the statistical quality function in white dwarf asteroseismology \citep[e.g.,][]{Giammichele2022}. 
Since we generally do not know the mode identifications ($\ell,k$) of the observed periods, each observed period ($P_\text{obs}$) is typically compared to the nearest model period ($P_\text{model}^{\text{nearest}}$) relative to the period uncertainty ($\sigma_{P_{\text{obs},i}}$) as
	\begin{equation}
	\chi^{2} = \sum_{i=1}^{n_{\text{obs}}} \frac{(P_\text{obs,\textit{i}} - P_\text{model, \textit{i}}^{\text{nearest}})^{2}}{\sigma_{P_{\text{obs},i}}^{2}},
    \label{eq:chi_sq_classic}
	\end{equation}
where $n_{\text{obs}}$ is the number of observed periods.
We will consider modifications to this quality function in later sections.
Occasionally other constraints on the spherical degree of the modes (e.g., from rotational splitting) are used to restrict fits to a subset of model modes with a specific $\ell$ value, but we allow our measurements to fit to either $\ell=1$ or 2 modes. Some fitting approaches ensure that no two measured periods are fitted to the same model mode \citep{Kim2007}, though our simulated period measurements are so far from each other that this is not relevant for the present analysis. 
Some works in the field characterize the fit residuals with other metrics, such as the root-mean-squared difference between measured and model periods, $\sigma_{\text{RMS}}$, with units of seconds \citep[e.g.,][]{Metcalfe2000}. While $\sigma_{\text{RMS}}$ conveys the typical scale of residuals in an intuitive way, $\chi^{2}$ is required in more sophisticated statistical analyses, such as error determinations or incorporating other data into the fitting.

Before we start fitting models to the data, we should adopt a criterion for what we will consider a statistically compelling fit worth further inspection. There will always be a best fit, but we should aim for our models to match the data at a level consistent with our understanding of uncertainty on both the measurements and modeling. While these uncertainties can be difficult to estimate accurately, we controlled the simulated noise in this parameter recovery experiment. We should expect to achieve a solution consistent with the correct underlying parameters (Table~\ref{tab:wdecparams}) with average $\sim 1\sigma$ residuals. It is tempting to adopt a strict significance criterion based on $\chi^2$ statistics, but the asteroseismic model fitting problem is not well defined for $\chi^2$ statistics \citep{Andrae2010}, as each measured pulsation period can be matched to numerous model pulsation modes. The interpretation of which discrete ``mode ID'' ($\ell,k$) corresponds to each observed period adds flexibility to the fitting, introducing complicated free parameters in addition to the varied parameters of mass and effective temperature. Even with three measured periods for a two-dimensional grid search, we can expect to find multiple degenerate solutions for different solutions to the mode identification problem. We will flag all models that fit within $\chi^{2}< \chi^{2}_{\rm threshold} = 10$ as a potential solution worth inspecting for this experiment. For fixed mode ID, we expect three measured periods for two free parameters to follow $\chi^2$ statistics with one degree of freedom, with a 0.15\% chance of the right solution having minimum $\chi^2 > 10$. With absolute magnitude included in the fit (two degrees of freedom), this increases to a 0.7\% chance of missing the correct solution. We validate in Section~\ref{subsec:errorestimation} that the solutions that meet this criterion reproduce all significant features in the probability distribution evaluated directly from the quality function. While using a smaller $\chi^{2}_{\rm threshold}$ could miss relevant solutions, using a more permissive threshold would flag additional solutions that would later be identified as having negligible contributions to the overall solution space.

\subsection{Interpolating a Coarse Grid to Finer Resolution}
\label{subsec:interpolation_finer_grid}

Our first step toward finding asteroseismic solutions to our simulated data is to compute a coarse grid of stellar models spanning most of the DAV instability strip using WDEC. We compute pulsation periods for $1{,}476$ WDEC models with masses sampled on the range $0.50 M_{\odot} \leq M_\star \leq 0.85 M_{\odot}$ in steps of $0.01 M_{\odot}$, and effective temperature sampled on the range $10{,}600\,{\rm K} \leq T_{\text{eff}} \leq 12{,}600$\,K in steps of 50\,K. While this does not encompass the entire mass and temperature range of known DAV pulsators \citep{Corsico2019Review} that might be relevant to real stars, it covers the correct solution for the fiducial model and is sufficient to demonstrate the statistical considerations presented in this work. Core parameters were set to the same values given in Table~\ref{tab:wdecparams} for this two-dimensional parameter recovery test. Due to numerical noise in the modeling of the base of the convection zone, we smooth over the model periods as a function of effective temperature with cubic fits as described in Appendix~\ref{app:convection}. 

\begin{figure}
    \centering
    \par\medskip
    \includegraphics[width=0.45\textwidth]
    {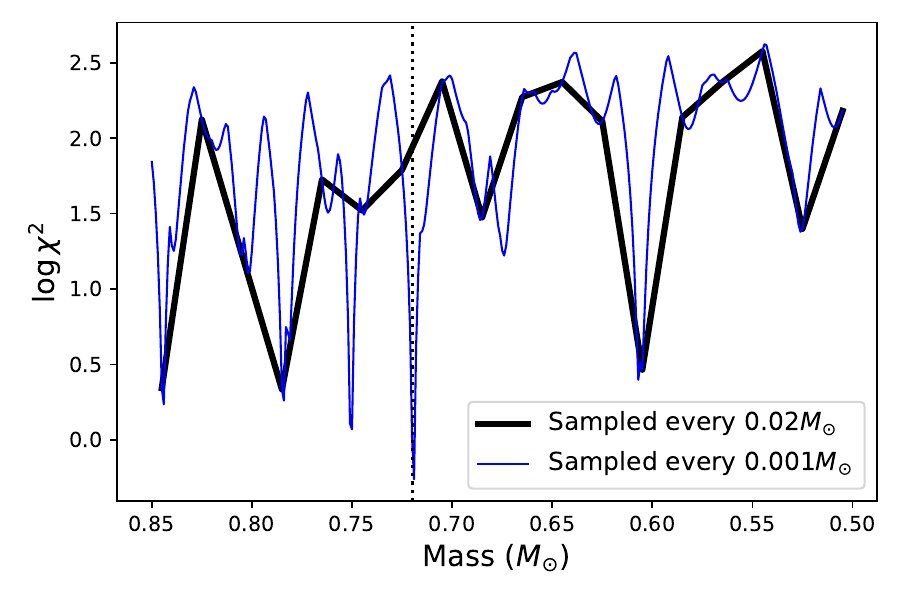}
    \caption{Comparison of $\chi^{2}$ goodness-of-fit curves for period grids of different mass resolution. The black curve corresponds to a coarse model grid that was sampled every $0.02 M_{\odot}$. The blue curve corresponds to a finely interpolated grid with periods sampled every $0.001 M_{\odot}$. The vertical dashed line (black) marks the true mass of the fiducial model. The fine grid resolves all solutions, while the coarse grid is insufficient and misses the correct solution.
    \label{fig:grid_comparison_mass}}
\end{figure}

With step sizes of $0.01M_{\odot}$ and 50\,K, our initial grid of WDEC models is too coarse to reliably resolve seismic solutions. To demonstrate this, we compare in Figure~\ref{fig:grid_comparison_mass} $\chi^{2}$ curves for our simulated period measurements as a function of mass (otherwise the same parameters of our true model) using both coarser ($0.02M_{\odot}$) and finer sampling ($0.001M_{\odot}$). The finer sampling is based on an interpolation of the period spectra, as detailed below. The choice in coarse step size is motivated by model resolutions used in representative studies in the literature \citep{Romero2012, Bell2019, Bognar2026}. While the finer sampling resolves individual local minima, the coarse sampling misses some of the degenerate solutions that are captured by the finer sampling \citep{Charpinet2015}. Most notably, the coarse grid misses the correct solution completely, demonstrating how valid solutions can fall through the cracks of a model grid with large step sizes.

\begin{figure*}
    \centering
    \par\medskip
    \includegraphics[width=0.8\textwidth,trim={0 1cm 0 2.5cm},clip]{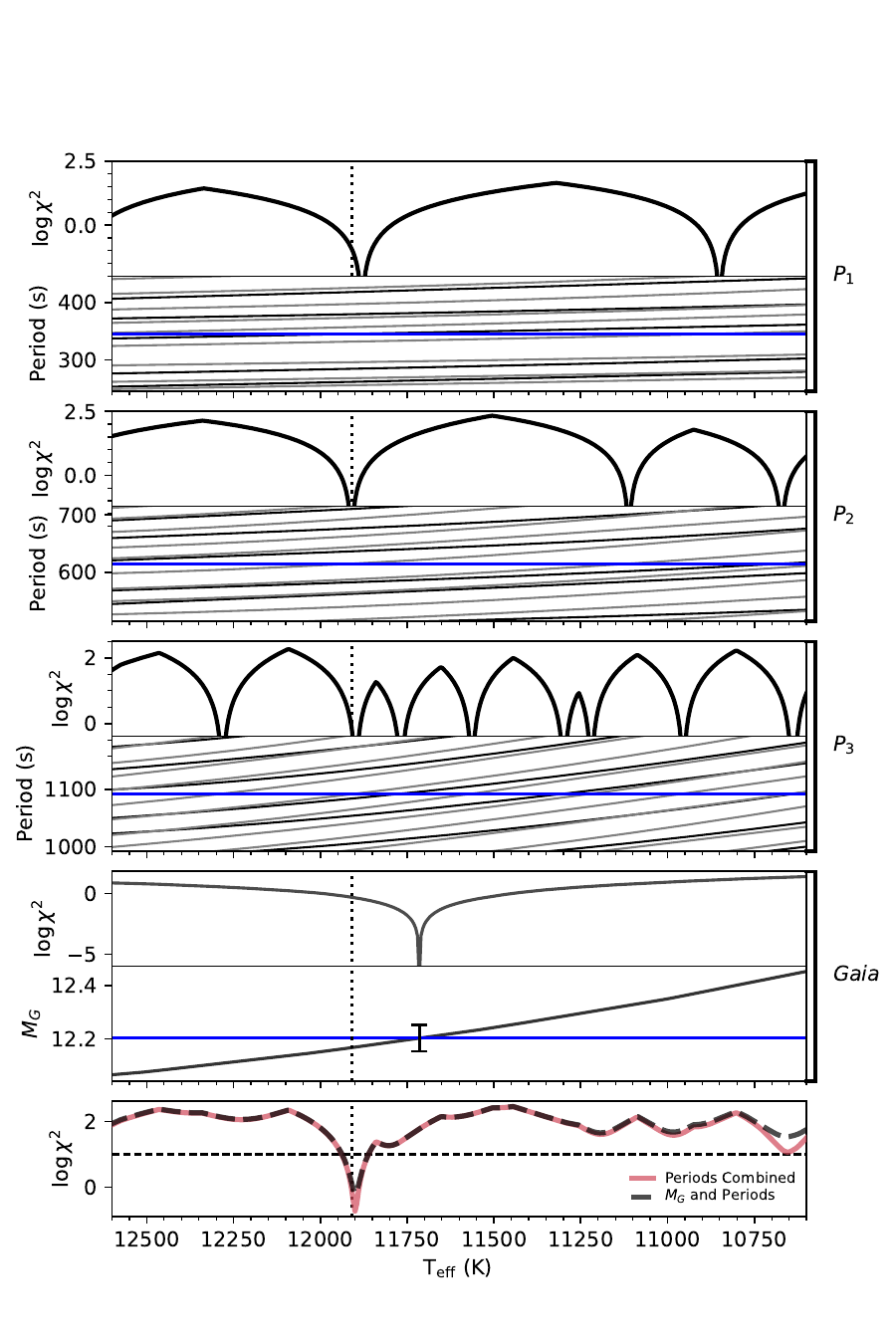}
    \caption{Comparison of simulated observations to an interpolated one-dimensional grid of pulsation periods and absolute magnitude as a function of effective temperature. All other stellar parameters are fixed to those displayed in Table~\ref{tab:wdecparams}. \textbf{The first three panels} show how each measured period $P_{1} = 345.5\pm1.0$\,s (panel 1), $P_{2} = 615.1\pm1.0$\,s (panel 2) and $P_{3} = 1092.1\pm1.0$\,s (panel 3) compares to the model periods. Evaluations of each period's contribution to the $\chi^2$ quality function (Eq.~\ref{eq:chi_sq_classic}) are shown above depictions of how dipole (black) and quadrupole (grey) pulsation periods come in and out of agreement with the measured periods (blue horizontal lines) at different effective temperatures. \textbf{The fourth panel} compares the simulated $G$-band absolute magnitude measurement (blue line with error bar represents the measurement uncertainty) to the modeled magnitude that varies monotonically with effective temperature. \textit{Gaia} absolute magnitude helps lift the solution degeneracy, as there exists a unique astrometric solution for fixed stellar mass. The bottom panel shows the combined $\chi^{2}$ goodness-of-fit curves versus effective temperature. \textbf{The thick red curve} considers all three periods together (Eq.~\ref{eq:chi_sq_classic}), and the 
    \textbf{dashed black curve} also incorporates the absolute magnitude constraint (Eq.~\ref{eq:chi_sq_gaia}). The $\chi^2_{\rm threshold} = 10$ constraint is shown with a horizontal dashed black line. The vertical dotted line is shown at the location of the effective temperature of the fiducial model.
}\label{fig:teff}
\end{figure*}

\begin{figure*}
    \centering
    \par\medskip 
    \includegraphics[width=0.8\textwidth,trim={0 1cm 0 2.5cm},clip]{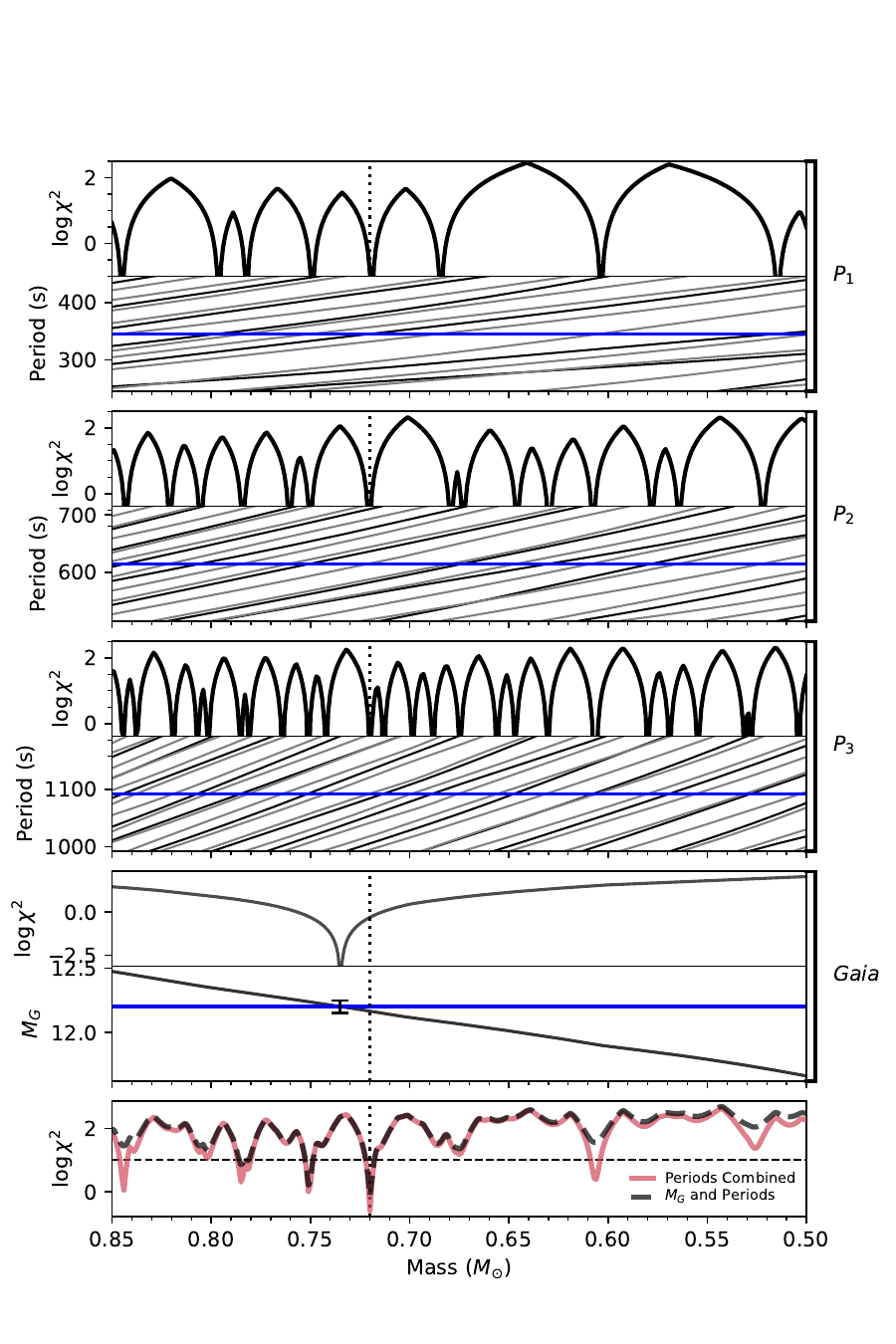}
    \caption{Same as Figure~\ref{fig:teff} but for a one-dimensional grid that samples stellar mass instead of effective temperature.
}\label{fig:mass}
\end{figure*}

To adequately resolve seismic solutions when varying model parameter $X$, the model grid should use step sizes ($\Delta X$) that produce changes in the model period ($P$) that are smaller than the period uncertainty ($\sigma_{P}$) over the range of the measured periods. We propose a criterion for sufficient grid resolution given by
\begin{equation}\label{eq:sampling}
    \Delta X < \frac{2 \sigma_P}{\max \left\{\left|\frac{dP}{dX}\right|\right\}},
\end{equation}
where the required resolution is set by the modes that are most sensitive to changes in this parameter. For each model period that can match to the measured periods, this ensures that there is at least one point on the grid where the model period will be sampled within $1\sigma$ of the measurement. From numerical estimates of $dP/dX$ for the range of observed periods across our grid, we estimate that reliably resolving solutions requires grid steps no larger than $\Delta M_{\star} = 0.001\,M_\odot$ and $\Delta T_{\rm eff} = 13.6$\,K.

Computing stellar structure models with this fine of resolution would be computationally expensive. However, as long as our coarse model grid captures the general response of pulsation periods to changes in model parameters, we can reliably interpolate pulsation periods onto a finer parameter grid of sufficient resolution to capture seismic solutions. We perform bilinear interpolation of the periods of each pulsation mode in our model grid, first in $T_{\text{eff}}$ in steps of 5\,K, then in $M_\star$ in steps of $0.001\,M_{\odot}$. This results in the well-resolved $\chi^2$ curve displayed in Figure~\ref{fig:grid_comparison_mass}.

Figure~\ref{fig:teff} shows the results of evaluating the $\chi^2$ quality function (Eq.~\ref{eq:chi_sq_classic}) comparing interpolated periods to the simulated period measurements as a function of effective temperature only. All other model parameters match those in Table~\ref{tab:wdecparams}. The top three subplots demonstrate how each of the three period measurements compare to the model periods. Each simulated period measurement is displayed as a horizontal blue line in the lower panels, and the interpolated model periods are displayed as black ($\ell=1$) and gray ($\ell=2$) curves. The periods are observed to generally increase with decreasing effective temperature. The upper panels display the logarithm of each period's contribution to the $\chi^2$ sum evaluated in the quality function (Eq.~\ref{eq:chi_sq_classic}). 
Each of the three observed periods intersects with different model mode periods at numerous values of $T_{\rm eff}$, producing sharp dips in their corresponding $\chi^2$ curves.
These intersections represent perfect agreement between measured and interpolated model periods, achieving an arbitrarily small contribution to $\chi^2$ depending on the size of the interpolation steps. 
The opportunity for a measured period to fit to any of the computed pulsation modes in the model grid is the source of solution degeneracy in asteroseismic fitting. While each measured period can be interpreted as matching to modes with different IDs ($\ell$ and $k$) at different $T_{\rm eff}$ values, these individual $\chi^2$ minima only coincide at the correct $T_{\rm eff}$ of our test model (vertical dotted line). The summed $\chi^2$ curve of the quality function in Eq.~\ref{eq:chi_sq_classic} for the three simulated period measurements is displayed as the red curve in the bottom subplot of Figure~\ref{fig:teff}. Only one minimum achieves a compelling value of $\chi^2< 10$ to be considered a decent fit to the data in this one-dimensional test, and it occurs at the correct value of $T_{\rm eff}$.

Figure~\ref{fig:mass} displays the results of a similar one-dimensional fit of the three simulated pulsation periods to our finely interpolated model grid, but as a function of stellar mass. White dwarf pulsation periods are highly sensitive to mass, with less massive stars with larger radii having longer pulsation periods. The individual measured period values come into agreement with different interpolated model periods at several mass values. In this case, the red curve representing the sum of $\chi^2$ terms from each of the three periods exhibits five compelling minima ($\chi^2< 10$) that should be considered degenerate candidate solutions for this data set.

Finally, we consider a two-dimensional parameter variation of both mass and effective temperature. Figure~\ref{fig:periods_only_contour_plots} panels 1--3 show how each of the observed periods come in and out of agreement with model periods for different combinations of mass and effective temperature. Each band in panels 1--3 corresponds to a track of agreement between a measured period and a different pulsation mode in the models defined by a specific spherical degree $\ell$ and radial order $k$. Intersections of these tracks across panels 1-3 produce the degenerate solution minima that are present in panel 4, representing the sum of $\chi^2$ terms from each observed period (Eq.~\ref{eq:chi_sq_classic}). Therefore, each degenerate solution is produced by a different candidate solution to the mode identification problem, i.e., which modes do we think we have observed? For this two-dimensional test with three measured periods, there are 28 combinations of model modes that produce values of $\chi^2 < 10$ that we consider to be compelling fits to the data. These solutions are only resolved and identified by the finely interpolated period grid.

\subsection{Lifting Solution Degeneracy with \textit{Gaia} Astrometry}

Seismic fitting generally permits degenerate solutions, and studies will often rely on post hoc constraints to select the most compelling solutions among these. One such constraint is the ``seismic distance'' \citep{Bell2019} based on a precision parallax measurement from \textit{Gaia}. In this section, we suggest that external constraints, such as those from astrometry, are better incorporated into the fitting in a statistically consistent manner by modifying the quality function. In this spirit, we suggest a modification to the $\chi^2$ quality function that takes into account the residual between the astrometric absolute magnitude measurement ($M_{G}\pm\sigma_{\text{mag}}$) and the absolute magnitudes simulated across the model grid ($M_{G,{\rm model}}$), given by
\begin{equation}
\chi^{2} =
\sum_{i=1}^{n_{\text{obs}}}
\frac{
\left(P_{\text{obs}, i} - P_{\text{model}, i}^{\text{nearest}}\right)^2
}{\sigma_{P_{\text{obs}, i}}^2
} + \frac{\left(M_{G} - M_{G,\text{model}}\right)^2}{
\sigma_{\text{mag}}^2}.
\label{eq:chi_sq_gaia}
\end{equation}
The model absolute magnitudes are calculated for WDEC models from Eq.~\ref{eq:interpolation} and interpolated onto the fine model grid by  bilinear interpolation as described for the model pulsation periods.

Panel 4 of Figure~\ref{fig:teff} illustrates how absolute magnitude varies monotonically with our one-dimensional variation of effective temperature. 
At fixed mass (approximately constant radius), a higher effective temperature translates to a greater luminosity and therefore a lower absolute magnitude; this comes into agreement with the measured absolute magnitude at a unique temperature in our 1D analysis at fixed mass. \textit{Gaia} absolute magnitude lifts the solution degeneracy, as there exists a unique astrometric solution (for fixed mass). Panel 5 of Figure~\ref{fig:teff} compares the combined $\chi^2$ curves that consider only the three measured periods (solid red curve; Eq.~\ref{eq:chi_sq_classic}) to that obtained when the absolute magnitude is included (dashed black curve; Eq.~\ref{eq:chi_sq_gaia}). The astrometric magnitude supports the global $\chi^2$ minimum at the correct solution (vertical dotted line) and rules out the secondary minimum near 10{,}650\,K that achieved $\chi^2$ near our adopted threshold $\chi^2_{\rm threshold} = 10$. 

\begin{figure}
	\centering
	\par\medskip
	 \includegraphics[width=0.4\textwidth]{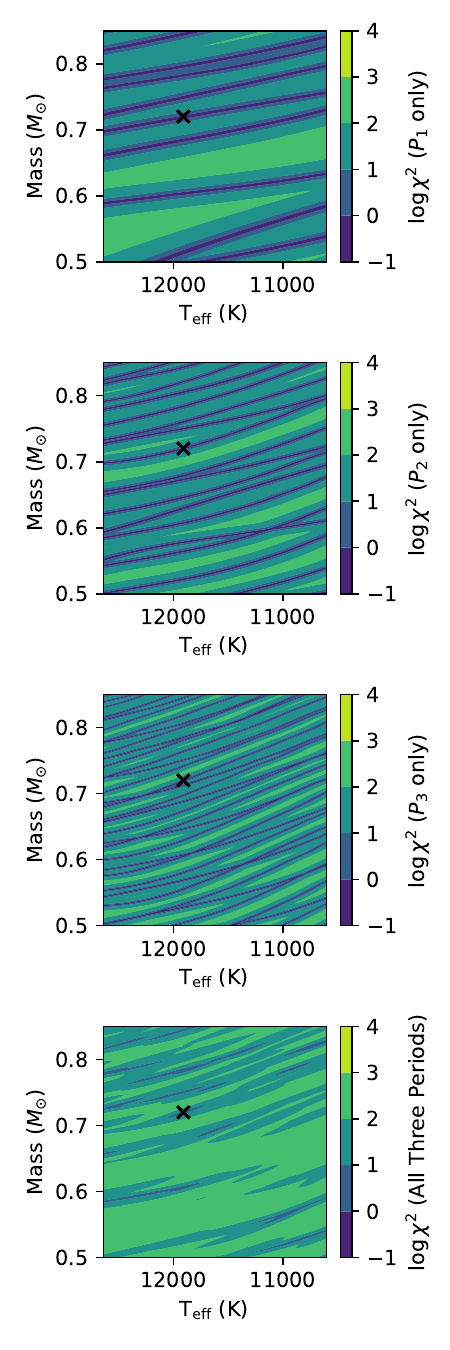}
     \caption{\textbf{Panels 1--3}: $\chi^{2}$ goodness-of-fit contours for each of the three simulated observed periods $P_{1}$ (345.54 s), $P_{2}$ (615.09 s), $P_{3}$ (1092.09 s) in the mass-effective temperature plane. \textbf{Panel 4}: $\chi^{2}$ contour plot for $P_{1}$, $P_{2}$, $P_{3}$ combined. The cross marks the location of the fiducial model.
     }
    \label{fig:periods_only_contour_plots}
     \end{figure}
     
\begin{figure}
	\centering
	\par\medskip
	 \includegraphics[width=0.4\textwidth]{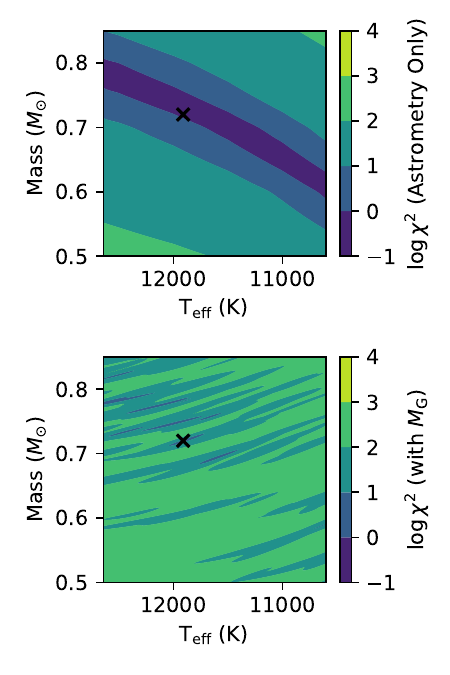}
    \caption{\textbf{Panel 1:} contour plot of the contribution to the $\chi^{2}$ quality function from only the astrometrically determined absolute $G$-band magnitude. \textbf{Panel 2}: $\chi^{2}$ contour plot for the combined fit, including absolute magnitude and pulsation periods in the quality function. \textit{Gaia} astrometry significantly reduces solution degeneracy, with the only viable solutions in blue.
    }
    \label{fig:Gaia_contour_plot}
    \end{figure}

\begin{figure*}
    \centering
    \par\medskip
    \includegraphics[width=0.8\textwidth]{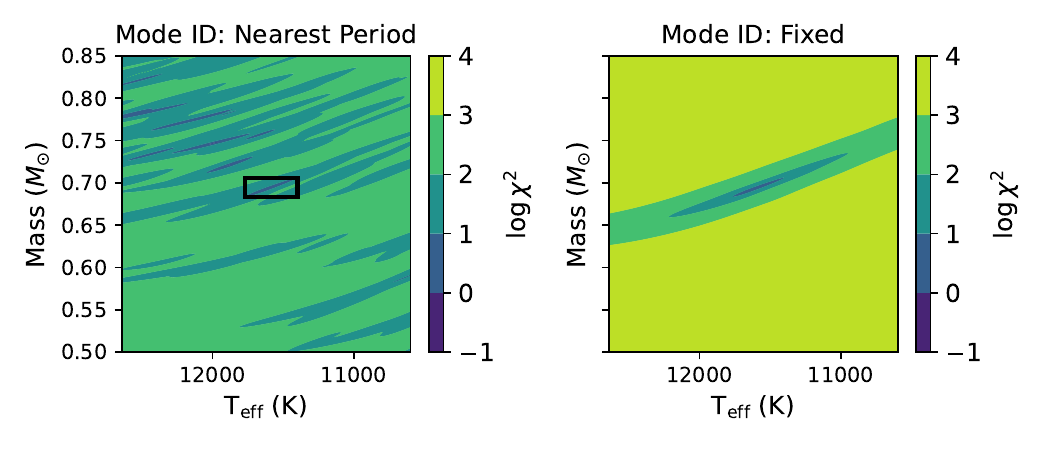}
    \caption{\textbf{Left Panel}: Same as panel 2 of Figure~\ref{fig:Gaia_contour_plot}, showing $\chi^2$ contours in Mass-$T_{\text{eff}}$ space from fitting absolute magnitude and periods to the nearest model periods (Eq.~\ref{eq:chi_sq_gaia}). A rectangular box highlights one of many degenerate solutions.   \textbf{Right Panel}: The highlighted solution is clearly represented by enforcing the mode IDs $(\ell, k)$ = (2,10), (2,21), (2,39) in the quality function (Eq.~\ref{eq:chi_sq_modeID}). }
\label{fig:mode_ID_comparison}
\end{figure*}

Similarly, panel 4 of Figure~\ref{fig:mass} shows that, for fixed effective temperature, absolute magnitude varies monotonically with mass. A less massive star has a greater luminosity and lower absolute magnitude because less massive white dwarfs generally have larger radii. Therefore, absolute $G$-band magnitude admits a unique solution at one specific mass for fixed effective temperature. 
Panel 5 compares the $\chi^2$ curves without considering absolute magnitude (red) and with absolute magnitude included (dashed black). 
\textit{Gaia} astrometry rules out at least two compelling secondary (degenerate) asteroseismic solutions purported by the red curve at masses above 0.83\,$M_{\odot}$ and below 0.62\,$M_{\odot}$. Including \textit{Gaia} astrometry reduces solution degeneracy, but it does not completely eradicate all distant degeneracies that satisfy the sensitivity threshold. 

The top panel of Figure~\ref{fig:Gaia_contour_plot} shows how an absolute magnitude determined from \textit{Gaia} astrometry admits a smooth, monotonic band of statistical agreement with the models in the two-dimensional space of mass and effective temperature. 
Compared to the degenerate solution space from considering only three pulsation periods shown in the bottom panel of Figure~\ref{fig:periods_only_contour_plots}, including the information from \textit{Gaia} (Eq.~\ref{eq:chi_sq_gaia}) reduces solution degeneracy across the grid, as shown in the bottom panel of Figure~\ref{fig:Gaia_contour_plot}. Compared to the 28 unique solutions with $\chi^2 < 10$ that resulted from considering periods only, there are only 11 compelling solutions with $\chi^2 < 10$ when the astrometric constraint is included. 

\subsection{Identifying Individual Solutions}

We assert that finding an asteroseismic solution is equivalent to finding a solution to the mode identification problem. Each solution ``island'' in the 2D parameter space (bottom panels of Figures~\ref{fig:periods_only_contour_plots} and \ref{fig:Gaia_contour_plot}) is produced where agreement between measured periods and individual model periods (top panels of Figure~\ref{fig:periods_only_contour_plots}) overlap. For a given $\chi^{2}$ minimum in parameter space, we seek to identify the $(l,k)$ values of the model modes that match to each of the observed periods. 
 
The left panel of Figure~\ref{fig:mode_ID_comparison} shows $\chi^2$ computed from Eq.~\ref{eq:chi_sq_gaia} (matching the bottom panel of Figure~\ref{fig:Gaia_contour_plot}), revealing a map of the degenerate solutions in model parameter space. One of the (incorrect) degenerate solutions is highlighted with a rectangle. We inspect the models near this candidate solution to identify the ($\ell$,$k$) values of the model modes nearest to the observed periods. If we modify the quality function to compare the measured periods to model periods fixed to these specific mode identifications, $P_\text{model,\textit{i}}^{\text{candidate}}$, as
\begin{equation}
	\chi^{2} = \sum_{i=1}^{n_{\text{obs}}} \frac{(P_\text{obs,\textit{i}} - P_\text{model,\textit{i}}^{\text{candidate}})^{2}}{\sigma_{P_{\text{obs}, i}}^{2}} + \frac{(M_{G} - M_\text{model})^{2}}{\sigma_\text{mag}^{2}},
    \label{eq:chi_sq_modeID}
	\end{equation}
a single solution emerges and all others disappear as seen in the right panel of Figure~\ref{fig:mode_ID_comparison}.
Isolating candidate solutions based on mode ID allows us to directly characterize and interpret them.
Repeating this procedure for each $\chi^{2}$ minimum allows us to uniquely identify all degenerate solutions.  

\begin{figure*}
    \centering
    \par\medskip 
    \includegraphics[width=0.95\textwidth]{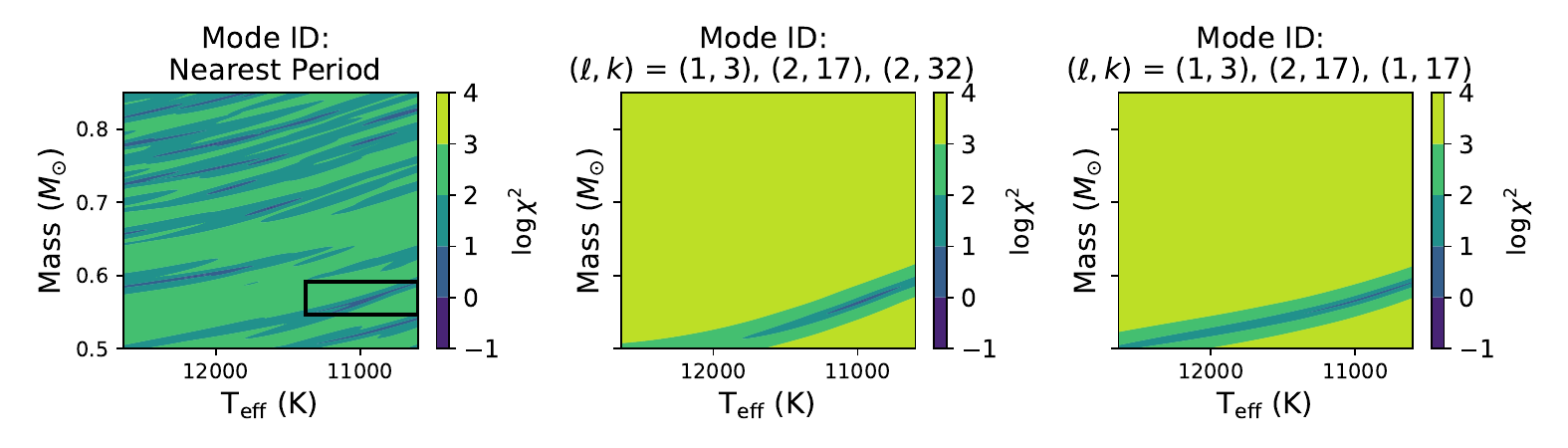}
    \caption{\textbf{Left Panel}: Same as panel 4 of Figure~\ref{fig:periods_only_contour_plots}, showing $\chi^2$ contours in Mass-$T_{\text{eff}}$ space from fitting only periods to the nearest model periods (Eq.~\ref{eq:chi_sq_classic}). A rectangular box highlights a composite island, where a measured period is in statistical agreement with more than one candidate pulsation mode. In this case, $P_3$ could be interpreted as either a (2,32) or a (1,17) mode. \textbf{Middle and Right Panels}: These overlapping degenerate solutions are disentangled using the fixed mode identification modification to the quality function (Eq.~\ref{eq:chi_sq_modeID}).}
\label{fig:composite_island}
\end{figure*}

\subsection{Beyond Nearest Periods}\label{subsec:beyondnearest}

By comparing the measurements to the nearest model periods in the quality function, there is potential to miss valid solutions. Considering the uncertainties on measured pulsation periods, there may be some locations in the grid where more than one model period is within statistical agreement to an observed period. In particular, in the plots of model period versus $T_{\rm eff}$ and $M_\star$ in Figures~\ref{fig:teff} and \ref{fig:mass}, there are values where curves representing $\ell=1$ and $\ell=2$ mode periods can be seen to overlap. Either one of these mode identifications could explain a period measured to have the value of the intercept, and if only the nearest model period to the measurement is considered, this could mask an equally valid solution. The equivalent issue is seen in our two-dimensional experiment where tracks representing different pulsation modes overlap in the top panels of Figure~\ref{fig:periods_only_contour_plots}.

To ensure that we capture all candidate solutions that fit within our adopted criterion of $\chi^2_{\rm threshold} = 10$, we extend the calculation of the quality function across the interpolated grid to report all combinations of mode IDs with periods that are close enough to satisfy this requirement, rather than only considering the nearest model periods. 
In order to identify viable combinations of mode IDs, we first record the mode IDs for all model periods that come within $\sqrt{\chi^2_{\rm threshold}}\sigma_{P}$ of each of our measured periods anywhere across the grid, as these residuals would not individually cause $\chi^2$ to exceed our threshold. Then we retain only those combinations of mode IDs that achieve a combined $\chi^2 < \chi^2_{\rm threshold}$ on the grid.
This provides us with a complete list of combinations of mode IDs that meet our fit criteria, defining all degenerate solutions of interest.
Table~\ref{tab:solutions} presents the mode ID combinations for all identified solutions with $\chi^{2}<10$ computed using Eq.~\ref{eq:chi_sq_modeID} when considering either only the periods or when also including absolute magnitude. 

Figure~\ref{fig:composite_island} demonstrates how two overlapping solutions that we detected in our periods-only analysis (bottom panel of Figure~\ref{fig:periods_only_contour_plots}) can be disentangled and inspected separately using specific mode IDs in the evaluation of the quality function (Eq.~\ref{eq:chi_sq_modeID}). The composite solution island highlighted with the black rectangle appears to split along two tracks in the left panel of Figure~\ref{fig:composite_island}, where only the $\chi^2$ value from the nearest model period is considered. Identifying all combinations of mode IDs that are good fits in this region reveals that $P_3$ could be interpreted to be either an ($\ell$,$k$) = (2,32) or (1,17) mode. The isolated solutions using these fixed mode identifications (Eq.~\ref{eq:chi_sq_modeID}) are shown in the second and third panels of Figure~\ref{fig:composite_island}, respectively. 

\begingroup
\begin{table*}
\centering
\hspace*{-3.2cm}
\resizebox{1.15\textwidth}{!}{
\begin{tabular}{|c||c|c|c|c|c|c||c|c|c|c|c|c|}
\hline
 &
\multicolumn{6}{c||}{Without Absolute Magnitude} &
\multicolumn{6}{c|}{With Absolute Magnitude} \\
\hline
$(\ell,k)$ Mode ID &
$\chi^2$ &
$M_{\star} \pm \sigma_{M_{\star}}  (M_\odot)$ &
$T_{\mathrm{eff}} \pm \sigma_{T_{\mathrm{eff}}} (K)$ &
Amplitude of Fit &
Correlation Coefficient &
Relative Likelihood &
$\chi^2$ &
$M_{\star} \pm \sigma_{M_{\star}} (M_\odot)$ &
$T_{\mathrm{eff}} \pm \sigma_{T_{\mathrm{eff}}} (K)$ &
Amplitude of Fit &
Correlation Coefficient &
Relative Likelihood \\ \hline

(2,11), (1,11), (1,23) & 0.245 & 0.7479 $ \pm $ 0.0043 & 11970 $ \pm $ 99 & 0.889 & $-$0.988 & 0.0626 & 0.358 & 0.7472 $ \pm $ 0.0037 & 11987 $ \pm $ 87 & 0.835 & $-$0.985 & 0.2108 \\

(1,5), (1,12), (1,25) & 1.712 & 0.7799 $ \pm $ 0.0070 & 12567 $ \pm $ 198 & 0.411 & $-$0.996 & 0.0580 & 1.758 & 0.7794 $ \pm $ 0.0056 & 12581 $ \pm $ 161 & 0.413 & $-$0.993 & 0.1917 \\

(1,5), (1,12), (2,46) & 0.841 & 0.7782 $ \pm $ 0.0034 & 12642 $ \pm $ 77 & 0.680 & $-$0.979 & 0.0400 & 0.870 & 0.7785 $ \pm $ 0.0033 & 12636 $ \pm $ 72 & 0.660 & $-$0.977 & 0.1491 \\

(2,12), (2,24), (1,24) & 0.165 & 0.7781 $ \pm $ 0.0052 & 12052 $ \pm $ 120 & 0.924 & $-$0.992 & 0.0817 & 1.852 & 0.7743 $ \pm $ 0.0047 & 12141 $ \pm $ 114 & 0.406 & $-$0.990 & 0.1336 \\

(1,4), (2,22), (2,41) & $\mathbf{0.192}$ & $\mathbf{0.7204 \pm 0.0033}$ & $\mathbf{11892 \pm 58}$ & $\mathbf{0.918}$ & $\mathbf{-0.977}$ & $\mathbf{0.0412}$ & $\mathbf{0.533}$ & $\mathbf{0.7211 \pm 0.0032}$ & $\mathbf{11880 \pm 54}$ & $\mathbf{0.772}$ & $\mathbf{-0.974}$ & $\mathbf{0.1323}$ \\

(2,10), (2,21), (2,39) & 0.103 & 0.6928 $ \pm $ 0.0039 & 11619 $ \pm $ 64 & 0.955 & $-$0.983 & 0.0477 & 0.878 & 0.6940 $ \pm $ 0.0038 & 11599 $ \pm $ 62 & 0.645 & $-$0.982 & 0.1256 \\

(2,11), (1,11), (2,43) & 1.876 & 0.7315 $ \pm $ 0.0040 & 12492 $ \pm $ 88 & 0.391 & $-$0.984 & 0.0267 & 5.036 & 0.7343 $ \pm $ 0.0035 & 12430 $ \pm $ 78 & 0.081 & $-$0.981 & 0.0191 \\

(2,11), (1,11), (2,42) & 0.036 & 0.7588 $ \pm $ 0.0027 & 11635 $ \pm $ 48 & 0.990 & $-$0.971 & 0.0332 & 4.204 & 0.7573 $ \pm $ 0.0026 & 11662 $ \pm $ 47 & 0.125 & $-$0.970 & 0.0169 \\

(2,13), (1,13), (2,48) & 0.310 & 0.8256 $ \pm $ 0.0036 & 12427 $ \pm $ 78 & 0.851 & $-$0.982 & 0.0487 & 5.890 & 0.8216 $ \pm $ 0.0040 & 12514 $ \pm $ 90 & 0.053 & $-$0.986 & 0.0142 \\

(1,3), (2,17), (1,17) & 4.954 & 0.5721 $ \pm $ 0.0096 & 10892 $ \pm $ 157 & 0.084 & $-$0.997 & 0.0111 & 8.667 & 0.5879 $ \pm $ 0.0064 & 10644 $ \pm $ 96 & 0.013 & $-$0.994 & 0.0040 \\

(2,12), (2,24), (2,44) & 2.666 & 0.7861 $ \pm $ 0.0038 & 11821 $ \pm $ 71 & 0.268 & $-$0.986 & 0.0131 & 8.536 & 0.7827 $ \pm $ 0.0033 & 11885 $ \pm $ 62 & 0.014 & $-$0.981 & 0.0027 \\

\hline 

(2,9), (2,19), (1,19) & 0.237 & 0.5980 $ \pm $ 0.0058 & 12177 $ \pm $ 186 & 0.887 & $-$0.993 & 0.1229  & \multicolumn{6}{c|}{---} \\

(1,3), (2,17), (2,32) & 0.332 & 0.5648 $ \pm $ 0.0058 & 10970 $ \pm $ 77 & 0.838 & $-$0.991 & 0.0532  & \multicolumn{6}{c|}{---} \\

(2,12), (2,23), (2,42) & 0.013 & 0.8247 $ \pm $ 0.0048 & 10634 $ \pm $ 77 & 1.018 & $-$0.992 & 0.0523  & \multicolumn{6}{c|}{---} \\

(2,10), (2,21), (1,21) & 4.567 & 0.6478 $ \pm $ 0.0055 & 12902 $ \pm $ 156 & 0.417 & $-$0.991 & 0.0520  & \multicolumn{6}{c|}{---} \\

(2,8), (2,16), (1,16) & 1.428 & 0.5306 $ \pm $ 0.0074 & 10784 $ \pm $ 136 & 0.497 & $-$0.996 & 0.0511  & \multicolumn{6}{c|}{---} \\

(2,13), (1,13), (1,26) & 0.707 & 0.8409 $ \pm $ 0.0048 & 11981 $ \pm $ 97 & 0.703 & $-$0.991 & 0.0487  & \multicolumn{6}{c|}{---} \\

(2,11), (1,11), (2,41) & 0.043 & 0.7855 $ \pm $ 0.0034 & 10918 $ \pm $ 57 & 0.999 & $-$0.981 & 0.0410  & \multicolumn{6}{c|}{---} \\

(1,5), (2,24), (1,24) & 1.251 & 0.8110 $ \pm $ 0.0041 & 11384 $ \pm $ 69 & 0.526 & $-$0.983 & 0.0297  & \multicolumn{6}{c|}{---}\\

(1,5), (2,24), (2,44) & 2.840 & 0.8118 $ \pm $ 0.0069 & 11383 $ \pm $ 120 & 0.264 & $-$0.996 & 0.0217  & \multicolumn{6}{c|}{---} \\

(1,4), (2,22), (2,42) & 9.320 & 0.6951 $ \pm $ 0.0019 & 12781 $ \pm $ 49 & 0.661 & $-$0.949 & 0.0217 
& \multicolumn{6}{c|}{---} \\

(2,12), (2,23), (1,23) & 1.275 & 0.8152 $ \pm $ 0.0037 & 10831 $ \pm $ 57 & 0.519 & $-$0.985 & 0.0206  & \multicolumn{6}{c|}{---} \\

(2,9), (2,19), (2,36) & 2.528 & 0.6062 $ \pm $ 0.0025 & 11893 $ \pm $ 47 & 0.285 & $-$0.957 & 0.0107  & \multicolumn{6}{c|}{---} \\

(2,13), (1,13), (2,47) & 3.910 & 0.8527 $ \pm $ 0.0037 & 11672 $ \pm $ 64 & 0.184 & $-$0.985 & 0.0082  & \multicolumn{6}{c|}{---} \\

(2,10), (1,10), (2,37) & 9.146 & 0.7272 $ \pm $ 0.0038 & 10534 $ \pm $ 63 & 0.019 & $-$0.984 & 0.0009  & \multicolumn{6}{c|}{---} \\

(2,13), (2,26), (2,48) & 9.198 & 0.8414 $ \pm $ 0.0045 & 12151 $ \pm $ 82 & 0.010 & $-$0.988 & 0.0007  & \multicolumn{6}{c|}{---} \\

(2,11), (2,23), (2,43) & 9.732 & 0.7428 $ \pm $ 0.0043 & 12269 $ \pm $ 91 & 0.008 & $-$0.988 & 0.0005  & \multicolumn{6}{c|}{---} \\

(1,6), (2,27), (2,50) & 9.700 & 0.8505 $ \pm $ 0.0020 & 12529 $ \pm $ 60 & 0.007 & $-$0.962 & 0.0002  & \multicolumn{6}{c|}{---} \\

\hline 

\end{tabular}}
\caption{Minimum $\chi^{2}$ goodness-of-fit values, optimal parameters, uncertainties, amplitudes, correlation coefficients, and relative likelihoods from 2D Gaussian fits to solution islands satisfying $\chi^{2} < 10$. Each solution corresponds to an interpretation of which modes the observed periods correspond to, with the $\ell$ and $k$ identifications for $P_1$, $P_2$, and $P_3$ listed. 
On the left are solutions found when considering only pulsation period measurements, and on the right are the same solutions that maintain $\chi^{2} < 10$ after including absolute magnitude. The upper set of rows are candidate solutions that remain viable when including absolute magnitude and are sorted by decreasing relative likelihood (including $M_G$). The lower rows are only compelling when ignoring absolute magnitude, sorted by decreasing relative likelihood.
The correct (fiducial) solution is indicated in bold.} \label{tab:solutions}
\end{table*} 
\endgroup

\subsection{Determinations of Parameter Precision} \label{subsec:errorestimation}

Having captured all degenerate solutions that pass our quality criteria, and equipped with a method to isolate each individual solution based on mode IDs, we are ready to fully characterize each solution. For each individual solution with $\chi^2$ calculated with fixed mode identification (Eq.~\ref{eq:chi_sq_modeID}), we convert our $\chi^{2}$ values to relative likelihoods proportional to $\exp(-\frac{\chi^{2}}{2})$ \citep[as in][]{VanGrootel2013}.

In general, the probability distributions for mass and effective temperature for individual solutions appear approximately Gaussian (Figure~\ref{fig:mode_ID_comparison}, right panel), and we fit Gaussians to each to determine optimal parameters and uncertainties, similar to the methodology of \citet[]{Giammichele2018}. We fit the probability distribution for each isolated solution ``island" with a rotated two-dimensional Gaussian with Astropy's \texttt{Gaussian2D} function. The parameters of the model represent the amplitude ($A$), central location coordinates ($T_{\rm eff},M_\star$), the rotation angle ($\theta$), and the scale parameters in the rotated frame ($\sigma_x$ and $\sigma_y$).  The fits converge from initial parameter values that we determine from a numerical estimate of the likelihood-weighted covariance matrix.

To get uncertainties on the parameters of mass and effective temperature, 
we must convert the fit parameters of the rotated two-dimensional Gaussian to model coordinates. The marginalized uncertainties on mass and effective temperature are
\begin{equation}\label{eq:marginalerrors}
\begin{aligned}
\sigma_{T_{\rm eff}} &= \sqrt{\sigma_x^2\cos^2{\theta} + \sigma_y^2\sin^2{\theta}}\\
\sigma_{M_\star} &= \sqrt{\sigma_x^2\sin^2{\theta} + \sigma_y^2\cos^2{\theta}}.
\end{aligned}
\end{equation}
These constraints are not independent, as evidenced by the tilt angle of the probability distributions. We report the correlation coefficients for each solution as 
\begin{equation}
    \rho = \frac{(\sigma_x^2 - \sigma_y^2)\sin{2\theta}}{2\sigma_{T_{\rm eff}}\sigma_{M_\star}}.
\end{equation}
Our full characterization of each solution from our two-dimensional tests, with and without the inclusion of absolute magnitude in the fitting, is reported in Table~\ref{tab:solutions}. 

\begin{figure*}
	\centering
	\par\medskip
	 \includegraphics[width=0.8\textwidth]{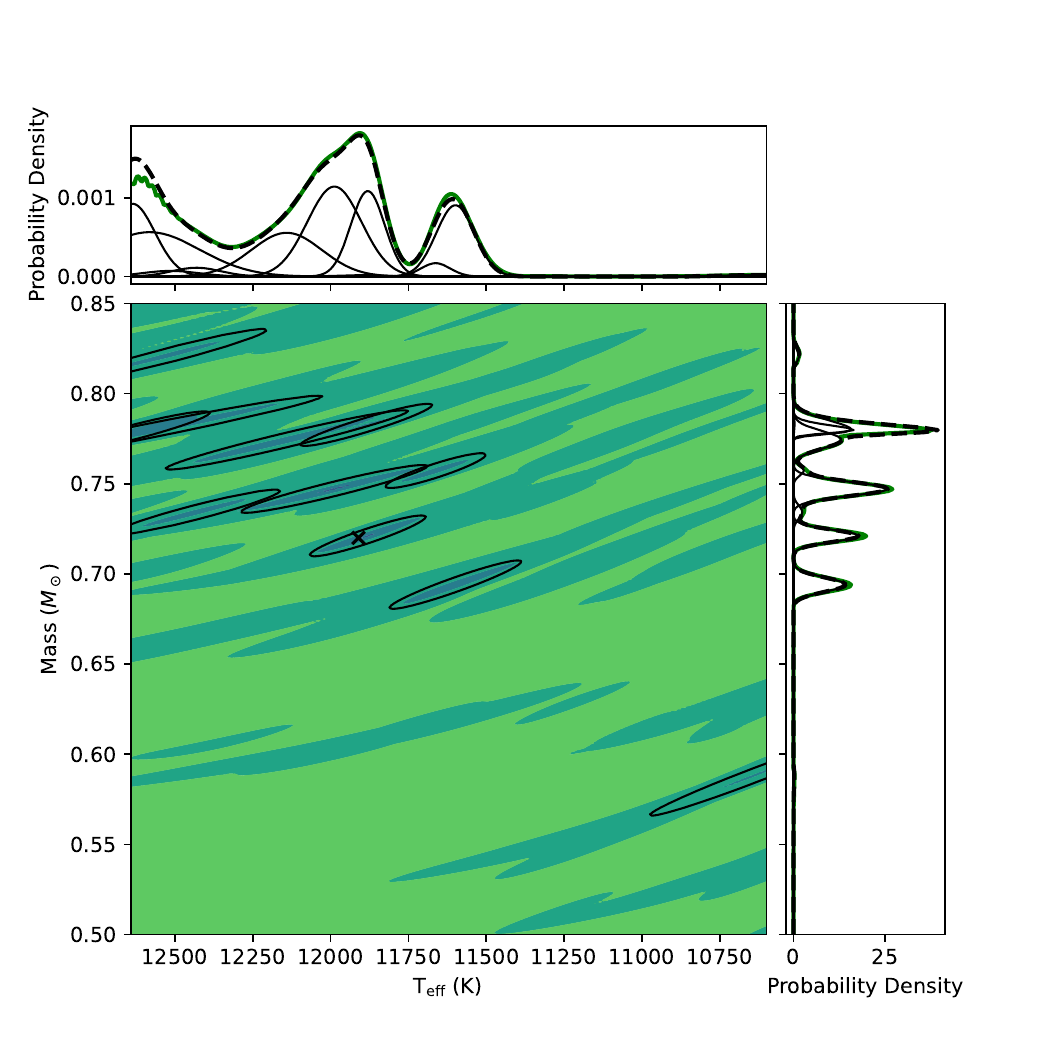}
	 \caption{Probability density functions (in black) for each of the islands containing the solutions of minimal $\chi^{2}$, shown out to 3$\sigma$. The dashed black line represents the combined (true) probability distribution for the system, both for mass and effective temperature. For comparison, the solid green line represents the probability distribution for both mass and effective temperature after marginalizing the likelihoods in either dimension. Here absolute magnitude is taken into account in the statistical fitting.}
    \label{fig:marginalized}
\end{figure*}

Finally, we assess the relative likelihood of each degenerate solution matching our measurements. The relative likelihoods of the solutions are treated as being proportional to the volume under each of the fitted Gaussians ($\propto A\sigma_x\sigma_y$). Assuming that one of the characterized solutions is the correct model, we normalize the relative likelihoods reported in Table~\ref{tab:solutions} to sum to unity. Of the 11 candidate solutions that obtained $\chi^2<10$ when including absolute magnitude, six have $>12\%$ relative likelihood individually, while the remaining five have $<2\%$ likelihood each. Of the six most compelling solutions, no individual solution stands out as clearly the most compelling solution for the star. The correct solution has the fifth highest relative likelihood. 
Figure~\ref{fig:marginalized} draws the $3\sigma$ contours from the Gaussian fits to each solution as black ovals around each local $\chi^2$ minimum on the contour plots of $\log{\chi^2}$ with absolute magnitude included in the fits (same as the bottom panel of Figure~\ref{fig:Gaia_contour_plot}). The top and right panels display marginalized distributions obtained by integrating over either mass or effective temperature. The individual Gaussian curves represent the individual solutions, and the dashed black curve shows the sum of these, representing the combined probability density curves for the parameters of mass and effective temperature as constrained by our models. The green curve shows the probability density function obtained by marginalizing the likelihoods directly in each dimension. Since this $\chi^2$ landscape considers only the nearest model periods to the measurements, it undercounts the ways that the models are compatible with the data where degenerate solutions overlap. We searched our interpolated period grid for all combinations of mode IDs that fit below our $\chi^2_{\rm threshold}$. The solutions that fit well to the measured periods having $(\ell,k)$ = \{(1,5), (1,12), (1,25)\}  and \{(1,5), (1,12), (2,46)\} overlap near $M_\star\approx0.77\,{\rm M}_\odot$ and $T_{\rm eff} = 12{,}500$\,K. We detect every viable solution individually, and identifying both of these solutions accounts for the additional probability density at high temperature compared to that implied by the nearest-period approach. Otherwise, our Gaussian fits are a striking match to the marginalized probability distribution achieved numerically with the nearest-periods metric. This supports that our method of searching for all combinations of mode IDs that cross $\chi^2_{\rm threshold}=10$ in the interpolated grid (Section~\ref{subsec:beyondnearest}) successfully captured all solutions with significant probability, and that the Gaussian fits faithfully characterize these solutions.

\section{Discussion and Conclusion} \label{sec: Section 4} 

In this work, we have demonstrated five strategies that allow us to identify and characterize the degenerate asteroseismic solution space from white dwarf period-by-period fitting. We present a parameter recovery test with three simulated pulsation periods and an absolute $G$-band magnitude computed for a structural white dwarf model computed with WDEC with parameters fixed to those in Table~\ref{tab:wdecparams}. With three measured periods, there are insufficient constraints to guide us to a unique solution. We compare the simulated measurements to both one- and two-dimensional grids of WDEC models with varying mass and/or effective temperature. 

First, we have demonstrated the utility of interpolating model pulsation periods for resolving individual solutions.  As long as the initial grid is sufficiently sampled to resolve the response of periods to structural changes in the star, periods can be interpolated onto a finer grid where changes in pulsation periods between adjacent grid points are smaller than the measurement uncertainties. This is necessary to ensure that the relevant solutions are not missed (Figure~\ref{fig:grid_comparison_mass}), and that uncertainties on solution parameters can be robustly determined.

Second, we show that incorporating constraints from \textit{Gaia} astrometry into the statistical fitting reduces solution degeneracy. Absolute magnitudes for our WDEC models were interpolated from synthetic photometry based on stellar atmosphere models (Equation~\ref{eq:interpolation}). In both one-dimensional (Figures~\ref{fig:teff} and \ref{fig:mass}) and two-dimensional (Figure~\ref{fig:Gaia_contour_plot}) parameter variation experiments, \textit{Gaia} astrometry supports the validity of the correct solution, while eliminating other solutions that would have been compelling if only the periods were considered. With simulated measurements of three pulsation periods, solution degeneracy in the two-dimensional test was significantly reduced by the inclusion of absolute magnitude in the fitting (Table~\ref{tab:solutions}), though the correct solution is still just one of multiple viable interpretations of the data. We note that the correct solution does not achieve the smallest $\chi^{2}$ value, so minimum $\chi^{2}$ is not a reliable metric for identifying the correct solution, even with a well-resolved model grid. Including the absolute magnitude alongside the measured periods in the $\chi^2$ quality function (Equation~\ref{eq:chi_sq_gaia}) treats these observables in a self-consistent manner for robust inference of stellar parameters; this is preferable to considering a ``seismic distance'' and a post-hoc consideration for selecting between multiple degenerate solutions \citep[e.g.][]{Bell2019, Uzundag2023, Bognar2026}. 

Third, we characterize the nature of each degenerate seismic solution as a different solution to the mode identification problem. That is, what are the $\ell$ and $k$ of each of the observed modes? For each $\chi^{2}$ minimum, 
and for each of the three observed periods, there exists a unique mode identification $(\ell, k)$ of the closest model period nearest the `observed' period. By modifying the quality function such that each of the three simulated period measurements is fit to the periods of specific model modes (Eq.~\ref{eq:chi_sq_modeID}), we can uniquely identify individual resolved solutions (Figure~\ref{fig:mode_ID_comparison}). When the model fitting produces many degenerate candidate solutions, this procedure can be repeated for each viable interpretation of the mode IDs being observed so as to resolve each and every solution. 

Fourth, we propose an extension to the use of the quality function to go beyond just considering the set of model periods nearest to the measured periods, as is often done in the literature \citep{Metcalfe2003b,Romero2012,Charpinet2015, Corsico2019, Uzundag2023}. To ensure that we do not miss viable solutions, we search the period grid for all combinations of periods that achieve statistical agreement with the measurements. This ensures that we capture all asteroseismic solutions that cross our adopted significance criterion of $\chi^2_{\rm threshold}=10$ that might otherwise be missed by the nearest-period approach. This effect can be seen in Fig.~\ref{fig:marginalized} in the combined probability distribution of our fitted solutions (dashed black curve) compared to the distribution obtained by directly marginalizing the results from the nearest-period metric (green curve). The curves coincide for nearly all temperatures and masses, except near a region of overlapping asteroseismic solutions at high temperature, where the nearest period approach misses the probability density associated with overlapping solutions.

Fifth, and finally, we have demonstrated a robust approach to determining reliable parameter uncertainties for each seismic solution. We fit two-dimensional Gaussian distributions to each of the isolated solutions obtained by fixing mode IDs in the computation of $\chi^2$. The parameter uncertainties can be obtained directly from the Gaussian fit parameters themselves. We find this to produce more reliable results than another approach that is often used to estimate uncertainties based on the work of \citet{ZhangRobinsonNather1986}. Our testing showed that applying the \citeauthor{ZhangRobinsonNather1986}\ approach to model grid fitting produces numerically noisy results, and that it has often been incorrectly implemented in the asteroseismology literature, as we describe in Appendix~\ref{app:Zhang}.

We have several advantages in this parameter recovery experiment using simulated data, and analyzing data for real pulsating white dwarf stars will require additional considerations. 
First, we are privy to the parameters of the model that we used to generate synthetic data, so we can be sure to recover the correct answer among our degenerate solutions. 
This allowed us to reduce dimensionality considerably by fixing model interiors to the correct structure, and typically interior parameters must be varied to asteroseismically solve the structures of real stars. We expect additional degeneracies to arise when more free parameters are varied in the models. Varying internal structure changes the effect of mode trapping in DAVs \citep{Winget1981}, where the response of pulsation periods is less linear than for the smooth monotonic changes observed when varying global parameters. Departures from monotonicity become salient in the presence of ``avoided crossings'' \citep{Aizenman1977} between periods of consecutive radial order. These phenomena are associated with non-linear variations in pulsation period with effective temperature, stellar mass, hygrogen envelope mass, and even crystallized mass fraction (in DAVs) \citep{Montgomery1999, Romero2012}. Linear interpolation over a model grid coarsely sampled in interior structure parameters may present limitations that will require additional models to be computed to resolve the physical period response.     

Real stars are not expected to perfectly match calculated model periods due to incomplete physics of the models. For example, there is inherent systematic uncertainty from numerical errors in modeling the location of the base of the convection zone in version 20 of WDEC, as illustrated in Appendix~\ref{app:convection}. This introduces systematic error, biasing the results to favor inaccurate models. One way to quantify the effects of such systematics would be to compare to results for an ensemble of objects obtained from independent methods \citep[e.g., spectroscopy;][]{Calcaferro2024}.
Systematics on absolute magnitudes from astrometric instrument calibration \citep{Lindegren2021} and extinction \citep{Vergely2022} must also be considered for real data sets, with errors propagated from astrometric parallax and apparent magnitude.

Our simulated data included Gaussian noise of 1.0\,s on each of the three simulated periods to match the scale of expected systematic errors in structural DAV models accounted for by \citet{Giammichele2022}. Typical residuals from grid fitting in the literature often exceed 1.0\,s, indicating either the presence of larger modeling systematics or the deficiencies of coarse grids that do not resolve solution space (Figure~\ref{fig:grid_comparison_mass}). We would expect that using larger error budgets in the analysis would result in additional degenerate solutions.

Finally, we only considered axisymmetric modes with $m=0$, and observed modes can be any components of multiplets typically interpreted as rotational splitting \citep[e.g.,][]{Bognar2024}. Our methodology is expected to identify additional degenerate solutions when measured periods can be fit to $m\neq0$ modes. For some pulsating stars, degeneracy can be reduced when definite $\ell$ and $m$ determinations are determined from observations of complete multiplets \citep[e.g.,][]{Hermes2017b} or fits to nonlinear pulse shapes \citep{Montgomery10, Provencal12}. The improvements demonstrated in our statistical framework for fitting pulsating white dwarf observations will allow us to engage with these challenges in pursuit of reliable asteroseismic results for real stars in the \textit{Gaia} era.

\vspace{5mm}

\begin{acknowledgments}
This material is based upon work supported by the National Science Foundation under Award AST-2406917. 
This project made use of computational systems and network services at the American Museum of Natural History supported by the National Science Foundation via Campus Cyberinfrastructure Grant Awards \#1827153 (CC* Networking Infrastructure: High Performance Research Data Infrastructure at the American Museum of Natural History) and \#1925590 (CC* Compute: High Performance Campus Computing for Institutional Research at the American Museum of Natural History).
Thanks to Barbara Castanheira for helpful discussions about this work. 
\end{acknowledgments}

\appendix

\section{Smoothing Over Convection Artifacts from WDEC}\label{app:convection}

\begin{figure}
	\centering
	\par\medskip
	 \includegraphics[width=0.95\textwidth]{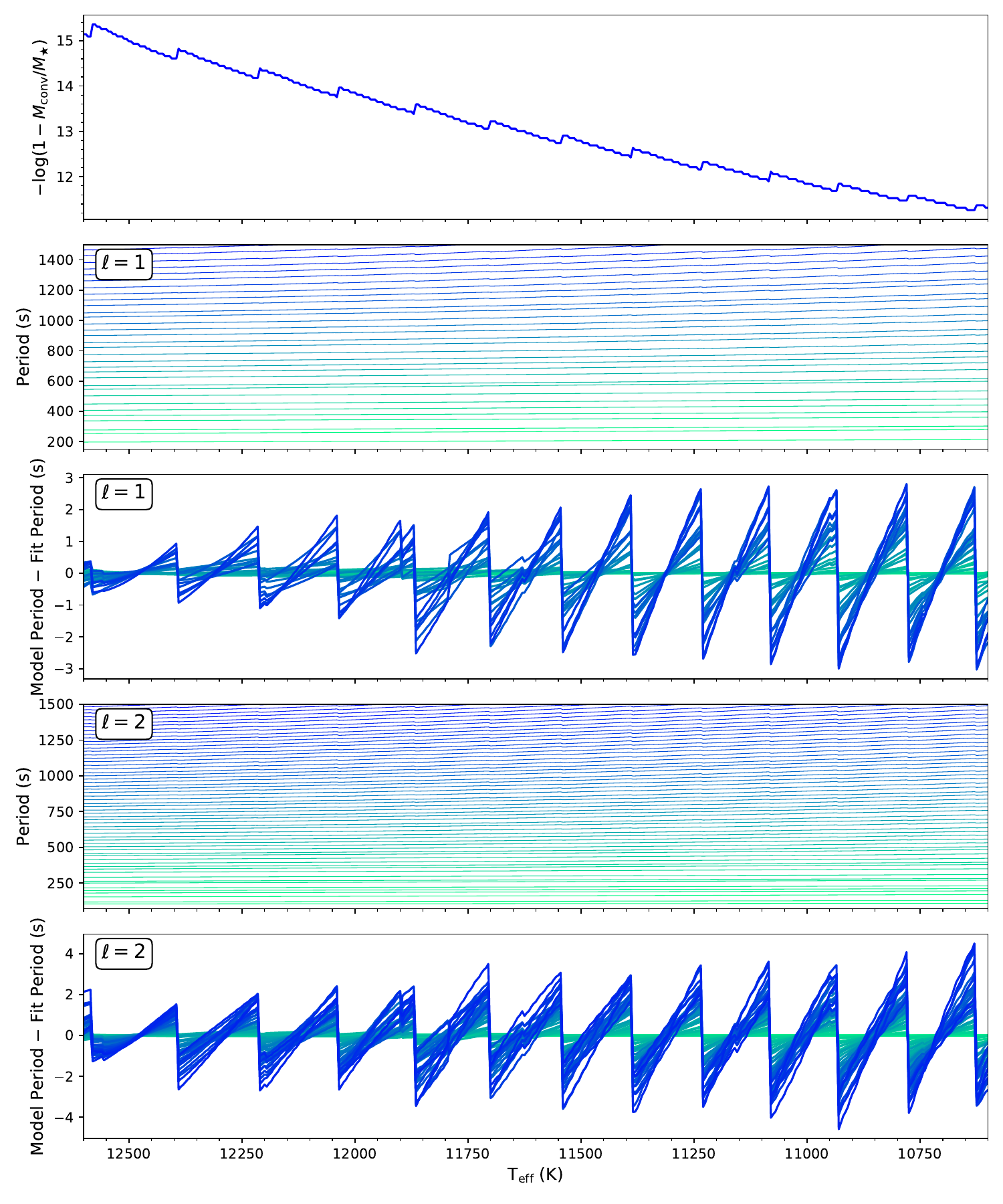}
	 \caption{\textbf{Panel 1}: Convection zone depth calculated in WDEC v20 as a function of effective temperature (other parameters fixed to values in Table~\ref{tab:wdecparams}). \textbf{Panel 2}: $\ell=1$ pulsation period spectrum vs.\ effective temperature from these models. \textbf{Panel 3}: Residuals after fitting cubic polynomials to the $\ell=1$ periods vs.\ effective temperature. \textbf{Panel 4}: $\ell=2$ pulsation period spectrum vs.\ effective temperature. \textbf{Panel 5}: Residuals after fitting cubic polynomials to the $\ell=2$ periods. The modes are colored by radial order to match to the corresponding residuals.
	 } 
\label{fig:appendix-cz_zone}
\end{figure}

We identified a numerical error in the modeling of the base of the convection zone in WDEC version 20 that produces irregularities in the period spectra. The top panel of Figure~\ref{fig:appendix-cz_zone} shows the depth of the convection zone in WDEC models with effective temperatures spanning the DAV instability strip in terms of $-\log{(1-M_{\rm conv}/M_\star)}$, where $M_{\rm conv}$ is the mass coordinate at the base of the convection zone. 
Other WDEC parameters, including MLT/$\alpha$ (convective efficiency), were fixed to the values given in Table~\ref{tab:wdecparams}.
While the convection zone generally deepens as the star cools, as expected, the apparent sawtooth appearance of the curve represents an unexpected systematic artifact in the modeling. At the same effective temperatures where the base of the convection zone jumps sharply in the models, subtle artifacts are present in the longest-period $\ell=1$ and $\ell=2$ modes. 
Panels 2 and 4 of Figure~\ref{fig:appendix-cz_zone} display the spectra of $\ell=1$ and $\ell=2$ mode periods computed by WDEC, which are dominated by the expected monotonic dependence on effective temperature. However, the residuals after fitting cubic polynomials to the periods of each mode that are displayed in panels 3 and 5 reveal jagged numerical artifacts related to the error in modeling the depth of the convection zone. The corresponding colors between panels reveal that longer period modes are more affected, as these are bounded by the base of the convection zone \citep{Montgomery2020}. The effect becomes more pronounced for lower effective temperatures and shorter periods, as the base of the convection zone recedes deeper into the star, affecting more modes. The jagged sawtooth behavior of the residuals represents numerical noise, causing systematic errors on the modeled pulsation periods of up to 3 seconds for $\ell=1$ modes and 4 seconds for $\ell=2$ modes at the cool edge of the instability strip. We use the periods evaluated from the cubic fits in the main analysis when performing our one- and two-dimensional grid parameter variation experiments in both mass and effective temperature. The cubic fits reproduce the overall monotonic trends of the periods while filtering out the sharp jumps caused by modeling artifacts, mitigating these systematics in this work. Further investigation is required to understand the source of these systematic modeling errors in version 20 of WDEC.

\section{Discussion of Zhang et al.\ Approach to Error Estimation}\label{app:Zhang}

The appendix of \citet{ZhangRobinsonNather1986} presents a framework for estimating uncertainties on best-fit parameters obtained from fitting synthetic light curve models to the eclipsing cataclysmic variable system HT Cassiopeiae. The authors assume a multi-normal probability distribution function in the parameter space of their models. Their method for estimating errors on model parameters is derived in their appendix section (d). The scale parameters of the multinormal distribution are estimated by measuring the rate with which $\chi^2$ increases with displacement in each dimension of parameter space from the optimal minimum. The specific steps required by their procedure to estimate the error on parameter $i$ can be summarized as follows:
\begin{enumerate}
\item Record as $S_0$ the minimum value of $\chi^2$ at the position in parameter space where $\chi^2$ is smallest.
\item Fix parameter $i$ to a value displaced by amount $d_{i}$ from $S_{0}$. 
\item Record as $S$ the minimum value of $\chi^2$ in this slice through parameter space, optimizing all other parameters besides $i$ to find the minimum $\chi^2$.
\item The error on parameter $i$ is recorded as $\sigma_{i}^{2} = d_{i}^{2}/(S - S_{0})$.
\end{enumerate}

Previous works in white dwarf asteroseismology have applied the \citeauthor{ZhangRobinsonNather1986}\ framework incorrectly, producing unreliable error estimates. Mistakes include:
\begin{itemize}
\item Applying this formula to $\chi^2$ values evaluated across a precomputed grid of models. In general, the location of the optimal $S_0$ is not present in a model grid, nor are the positions of $S$ that are offset in each dimension. Even for our interpolated grid that samples solutions within $1\sigma$ (Section~\ref{subsec:interpolation_finer_grid}), we found error estimates based on minima $S_0$ and $S$ located on the grid to be numerically noisy.
\item Applying this approach to a complex $\chi^2$ landscape evaluated for ``nearest periods'' (Eq.~\ref{eq:chi_sq_classic}) that does not resemble a multinormal distribution. Section~\ref{subsec:errorestimation} showed that seismic solutions evaluated for fixed mode IDs (Eq.~\ref{eq:chi_sq_modeID}) do resemble multinormal distributions and could be characterized with the \citeauthor{ZhangRobinsonNather1986}\ approach. If evaluated against the nearest model periods, however, taking a step in parameter space could move you from sampling $S_0$ for one seismic solution and $S$ for a different solution. This is especially a problem if the grid is coarsely sampled (see, e.g., Figure~\ref{fig:grid_comparison_mass}).
\item Some works have substituted other fit metrics for $\chi^2$, such as $\sigma^2_{\text{RMS}}$ \citep[with units of s$^2$;][]{Corsico2019,Uzundag2023}, or other incompatible metrics, such as the average of absolute differences \citep{Romero2012}, or amplitude-weighted $\sigma^2_{\text{RMS}}$ \citep{Castanheira2008,Romero2019}.
\end{itemize}

\begin{figure}
	\centering
	\par\medskip
	 \includegraphics[width=0.5\textwidth]{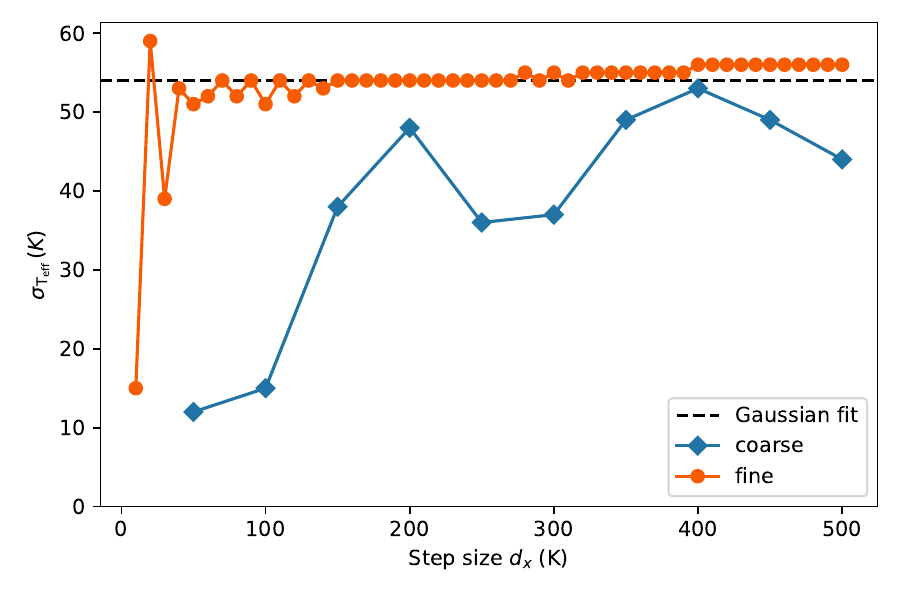}
	 \caption{Estimated error on effective temperature as a function of parameter step size, according to the prescription by \cite{ZhangRobinsonNather1986}, when applied to the correct solution fitted with fixed mode identification. The coarse model grid yields poor estimates for parameter uncertainty compared to the fine model grid, which converges to a value consistent with the error predicted by a multinormal fit.}
    \label{fig:Zhang_teff}
    \end{figure}

\begin{figure}
	\centering
	\par\medskip
	 \includegraphics[width=0.5\textwidth]{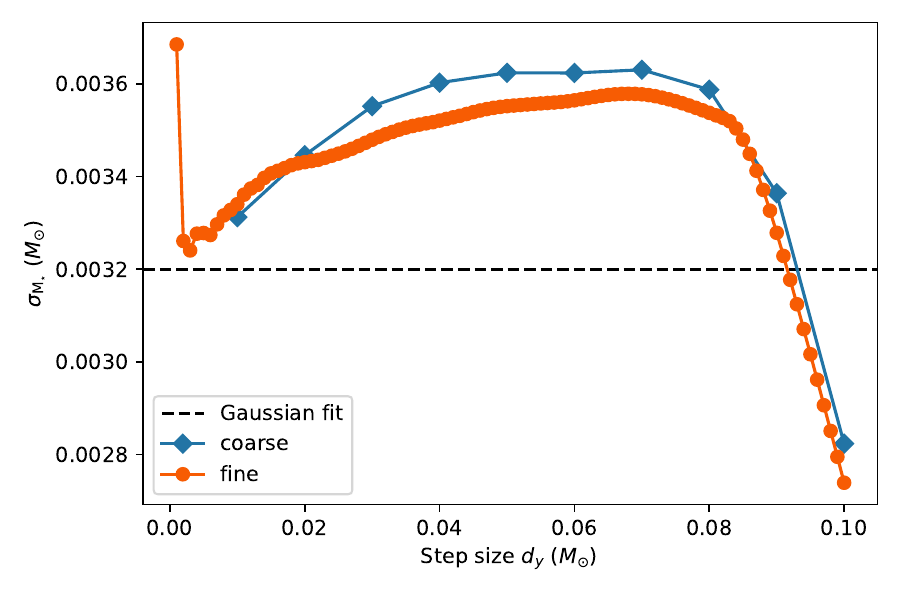}
	 \caption{Estimated error on mass as a function of parameter step size, according to the prescription by \cite{ZhangRobinsonNather1986}, when applied to the correct solution fitted with fixed mode identification. Neither the coarse nor fine model grids exhibit numerically stable error estimates.}
    \label{fig:Zhang_mass}
    \end{figure}

Figure~\ref{fig:Zhang_teff} and Figure~\ref{fig:Zhang_mass} test the numerical stability of error estimates obtained when applying the \cite{ZhangRobinsonNather1986} approach to the correct solution fitted with our model grids with fixed mode identification. 
Using a coarse grid, the estimated uncertainties following \citeauthor{ZhangRobinsonNather1986} are poor approximations compared to our Gaussian fitting results, with significant numerical noise that is sensitive to the choice of initial step size $d_i$.
The results improve when using the finely interpolated grid, and should converge to the correct answer with increased grid resolution.
This exposes the limitations and inaccuracies of applying this approach to model grid-fitting of white dwarfs, even though it is a mathematically sound and valid technique for error estimation in other contexts when applied as formulated by \citeauthor{ZhangRobinsonNather1986}. For fitting measurements to a model grid, we found uncertainties to be more reliably estimated from the parameters of a best-fit multinormal distribution to the likelihood function, as described in Section~\ref{subsec:errorestimation}.

\bibliography{bibliography}{}
\bibliographystyle{aasjournalv7}

\end{document}